\documentclass{CUP-JNL-EDS}

\usepackage{latexsym}
\usepackage{graphicx}
\usepackage{multicol,multirow}
\usepackage{amsmath,amssymb,amsfonts}
\usepackage{mathrsfs}
\usepackage{amsthm}
\usepackage{rotating}
\usepackage{appendix}
\usepackage[authoryear]{natbib}
\usepackage{ifpdf}
\usepackage[T1]{fontenc}
\usepackage{times}
\usepackage{newtxmath}
\usepackage{textcomp}%
\usepackage{xcolor}%
\usepackage{hyperref}
\usepackage{lipsum}
\usepackage{subcaption}
\usepackage{adjustbox}
\usepackage{tikz}
\usetikzlibrary{positioning, arrows.meta, calc, fit, shapes.geometric, shapes.misc, matrix, backgrounds}
\usepackage{pifont}

\newcommand{\cmark}{\ding{51}}
\newcommand{\xmark}{\ding{55}} 

\articletype{}
\jyear{2026}
\jvol{4}
\jdoi{10.1017/eds.2020.xx}

\begin{document}

\makeatletter
\def\@opjournalheader{} 
\def\@copyrighttext{} 

\renewcommand{\ps@titlepage}{%
     \def\@oddhead{} 
     \let\@evenhead\@oddhead%
     \def\@oddfoot{} 
     \let\@evenfoot\@oddfoot%
}
\makeatother

\begin{Frontmatter}

\title[Article Title]{A Fortran General-Purpose Transpiler: Proof of Concept}

\author[1]{Shivamshan Sivanesan}
\author*[1]{Kazem Ardaneh}\email{kardaneh@ipsl.fr}\orcid{0000-0003-0473-6907}
\address[1]{\orgdiv{Climate Modeling Center}, \orgname{Sorbonne University, CNRS, IPSL}, \orgaddress{\city{Paris}, \country{France}}}

\authormark{Sivanesan \& Ardaneh}

\keywords{Fortran, Python, Transpiler, Source-to-source compiler, High performance computing, Automatic differentiation}

\abstract{Fortran has been the cornerstone of high-performance computing for decades and remains unmatched in many domains. Yet the language faces an expertise gap: a new generation of scientists is barely familiar with it, while many experienced Fortran developers are only now transitioning to modern ecosystems such as JAX. This gap often results in "Fython"--Python code written with a Fortran mindset-- that fails to leverage modern frameworks. We present FGPT, a transpiler framework designed to bridge this gap. It provides a systematic pipeline that transpiles Fortran into GPU-adapted Fortran, auto-differentiable Fortran via Tapenade, or NumPy and JAX scripts. Its architecture comprises three stages: (i) a frontend that parses Fortran and extracts target procedures along with all their dependencies; (ii) a middle-end that lowers the code into an intermediate representation, then into GPU-adapted or auto-differentiable Fortran, or a NumPy class; and (iii) a backend that transforms NumPy scripts into JAX modules optimized for GPU acceleration and automatic differentiation. Large language models fail when applied to the scale of community scientific codes—often spanning hundreds of thousands of lines—where consistent transformations, strict numerical fidelity, and validation against production tests are non-negotiable. FGPT addresses these challenges by preserving program semantics throughout the entire translation. We verified the framework on representative climate modeling kernels and demonstrated that it produces correct, differentiable Python implementations without requiring manual intervention. By combining rigorous compiler techniques with modern accelerator support, FGPT offers a scalable, trustworthy path for modernizing legacy Fortran code.}


\end{Frontmatter}

\section[Introduction]{Introduction}
Fortran has long been the dominant language for high-performance computing (HPC), owing to its unrivaled efficiency in loop-oriented computations. Today, the majority of climate models, weather forecasting systems, computational fluid dynamics solvers, and other large-scale scientific codes are written in Fortran. Yet as these codes continue to grow in size and complexity, their compilation, maintenance, and evolution have become increasingly challenging.

Modernizing legacy Fortran presents a major hurdle. Adapting existing applications to heterogeneous architectures--including GPUs--often demands extensive manual refactoring, a task hampered by a shrinking pool of Fortran experts. This challenge is compounded by the coexistence of multiple language standards, compiler implementations, and project-specific coding conventions. Moreover, compared with high-level ecosystems such as Jax, Fortran offers limited native access to automatic differentiation (AD), AI-driven frameworks, and accelerator programming models. These limitations increasingly isolate Fortran codes from the broader computational science innovation cycle.

Modernization is therefore essential--not only to extend the lifetime of established models, but also to unlock new computational paradigms such as differentiable programming and hybrid CPU-GPU execution. Developing scalable, automated approaches for translating legacy Fortran applications is thus critical for the sustainability and future evolution of scientific code. Existing interoperability tools, such as \texttt{f2py} \citep{peterson_2009}, allow Python to call compiled Fortran code but do not generate equivalent source code—and thus do not meet the demand for fully translatable, maintainable Python implementations.

The primary approach to porting Fortran to higher-level languages remains manual reimplementation. \citet{hafner_veros_2018} demonstrated this with Veros, a JAX-based ocean circulation model developed through regular-expression preprocessing followed by extensive manual vectorization--a process that reproduced the original model's numerics with competitive GPU performance but required roughly one year of expert effort and yielded a methodology specific to a single application. More recently, \citet{zhou_proof_concept_2024} explored LLM-based translation of land-surface routines into Python/JAX using GPT-4 with dependency-ordered decomposition and iterative check against unit tests. While this achieved GPU acceleration and enabled gradient-based parameter calibration, the authors concluded that whole-code translation remains impractical: legacy scientific applications exceed current context limits, and function-level translations frequently require manual correction.

Agentic approaches have recently been applied to larger codes. \citet{lahlou_systematic_2026} translated the 19,000-line CLM-ml-v2 model into a numerically and gradient-verified JAX implementation using dependency analysis, persistent state documents, and a Fortran reference oracle. Similarly, \citet{koldunov_ocean_2026} reported a multi-week agent-assisted migration of the 74,000-line FESOM2 ocean model to C++/Kokkos. Despite their differing methodologies, both studies demonstrate a consistent conclusion: large, configurable applications demand incremental, dependency-aware decomposition over monolithic translation.

Existing approaches remain limited either by the extensive human effort required for manual translation or by the probabilistic nature of LLM-based workflows, which often demand iterative check, debugging, and human intervention--undermining reproducibility and independent verification. Moreover, no prior approach has demonstrated a deterministic transformation pipeline capable of scaling to large scientific codes while preserving numerical fidelity. Compiler-based transformations, in contrast, are deterministic, reproducible, and independently verifiable; each stage can be validated in isolation while naturally scaling to large applications. These facts motivate the following question:

\begin{quote}
\textit{Can large legacy Fortran codes be automatically modernized via a deterministic, dependency-aware transpiler pipeline that preserves numerical fidelity while targeting modern hybrid architectures?}
\end{quote}

We introduce FGPT (Fortran General-Purpose Transpiler), a source-to-source transpiler that combines dependency-aware procedure isolation with Abstract Syntax Tree (AST) based transformations to modernize legacy Fortran code.
FGPT parses the original code, isolates target procedures along with their dependencies, generates reference input–output pairs from the original code, and translates each isolated procedure into GPU-adapted Fortran, auto-differentiable Fortran via Tapenade \citep{hascoet_2013}, or Python. For the Python target, FGPT first emits a NumPy implementation, then transforms it into JAX, with numerical consistency verified against the original Fortran after each stage. The current work focuses on the Python translation pipeline. We evaluate FGPT on the largest procedures of the IPSL land-surface model, comprising interconnected modules with deeply nested procedure hierarchies exceeding 15,000 lines of Fortran code. We show that FGPT correctly reconstructs procedure interfaces and dependencies while producing numerically consistent NumPy and JAX programs. 

\section[Methodology]{Methodology}\label{sec:method}

\begin{figure}[t]
        \centering
        \begin{tikzpicture}[
            node distance=3cm and 3cm,
            scale = 0.6,
            transform shape, 
            >=Stealth,
            box/.style={
                draw,
                rectangle,
                minimum height=1cm,
                minimum width=3.2cm,
                align=center,
                font=\large,
                line width=0.6pt,
                rounded corners=4pt
            },
            fortran/.style={box, fill=yellow!20},
            ast/.style={box, fill=blue!15},
            python/.style={box, fill=green!15},
            explain/.style={box, fill=gray!10, minimum width=4.5cm},
            arrow/.style={->, line width=0.8pt},
            line/.style={-, line width=0.8pt},
            dottedarrow/.style={->, line width=0.6pt, dashed}
        ]
        
        \node[fortran] (fortran) 
        {\textbf{Fortran Code}\\
        Original source code};
        
        \node[ast, right=of fortran] (ast) 
        {\textbf{AST}\\
        Semantic representation};
        
        \node[python, right=of ast] (python) 
        {\textbf{Python Code}\\
        Generated Python syntax};
        
        \node[explain, below=2.5cm of ast] (intermediate) 
        {\textbf{Create an intermediary representation using the AST}};
        
        \node[ast, below= 6cm of fortran] (intent) 
        {\textbf{Encodes program intent}};
        
        \node[ast, below=6cm of ast] (numpy) 
        {\textbf{Leverages Python libraries}\\
        (e.g., NumPy)};
        
        \node[ast, below=6cm of python] (meaning) 
        {\textbf{Preserves semantic meaning}};
        
        \draw[arrow] (fortran) -- (ast);
        \draw[arrow] (ast) -- (python);
        
        \draw[arrow] (ast) -- (intermediate);
        
        \coordinate (merge) at ($(intermediate.south)+(0,-0.8cm)$);
        
        \draw[arrow] (merge) -- (intermediate.south);
        
        \draw[line] (intent.north)  |- (merge);
        \draw[line] (numpy.north)   |- (merge);
        \draw[line] (meaning.north) |- (merge);

        \draw[dottedarrow] 
        (ast) to[bend right=45] 
        node[midway, below, font=\bfseries]{One line of code: \texttt{ast.unparse(tree)}}
        (python);
        \end{tikzpicture}
        \caption{AST-based transpilation}
        \label{Fig1}
\end{figure}
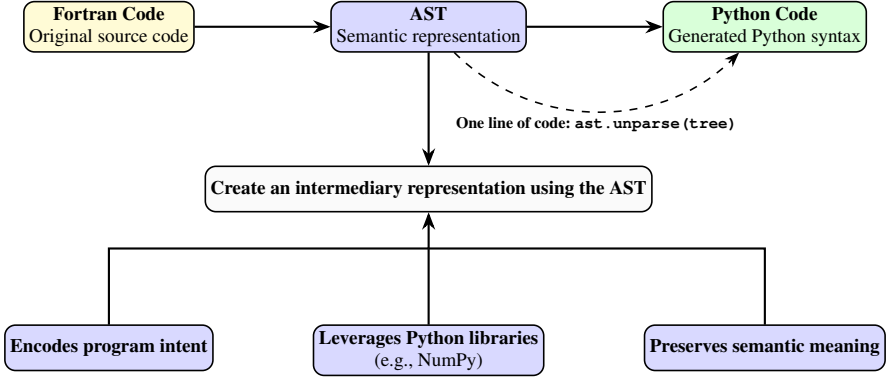

FGPT begins by parsing Fortran source with \texttt{fparser} \citep{{fparser_2026}} and representing it as an AST, then applies a sequence of deterministic semantic transformations for generating the final code (Figure~\ref{Fig1}).
For the transpilation, a line-by-line string translation strategy was rejected because Fortran and Python differ fundamentally in syntax, execution model, array indexing, loop semantics, memory layout, intrinsic procedures, and control-flow. Instead, FGPT adopts a compiler-inspired intermediate representation using the Python AST. Transformations operate on program constructs rather than source strings, enabling semantic-preserving rewrites independent of formatting or coding style. The AST is recursively traversed to identify procedures, loops, conditionals, intrinsics, declarations, array operations, procedure calls, etc. These nodes drive construction of the corresponding Python AST using Python's \texttt{NodeTransformer} framework, systematically rewriting Fortran-specific constructs into equivalent Python representations. Executable Python script is then regenerated via \texttt{ast.unparse}.

Using the AST as the intermediate representation offers three principal advantages: modularity (individual passes can be introduced or extended independently), semantic transformations on program structure rather than strings, and a foundation for subsequent passes, including NumPy to JAX  lowering. The same methodology extends to the JAX backend: rather than translating Fortran directly to JAX, the validated NumPy AST serves as the intermediate representation from which JAX code is generated through additional semantic transformations. This staged design enables independent validation of each transformation phase while progressively extending the capabilities of the pipeline.

Scientific Fortran applications often comprise thousands of procedures across numerous modules with complex interdependencies, making whole-code transpilation impractical due to the difficulty of analysis, transformation, and validation. To address this, FGPT employs a dependency-aware procedure isolation stage that decomposes the code into independently executable units prior to transpilation--establishing the foundation for all subsequent transformations, as detailed below.

\subsection{Isolation}\label{sec:isolation}
The objective of the isolation stage is to extract a Fortran procedure together with all information required for its independent compilation, execution, verification, and subsequent transpilation. This requires reconstructing its complete execution context--including variable declarations, array allocation semantics, intrinsic procedures, external calls, module variables, interface blocks, and other constructs upon which the procedure depends.

As each procedure is extracted, structural metadata--including declarations, dependencies, interfaces, and the call graph--is simultaneously collected and forwarded to the AST transformation pipeline. Thus, the isolation stage not only produces an independently executable unit for numerical validation but also provides the semantic context required for subsequent transformations. The extracted dependencies (called procedures and global variables) are assembled into a generated \texttt{global} module, while the isolated target procedure is embedded within a generated \texttt{main} program. The isolation process consists of six components:

\begin{itemize} 
\item \texttt{Processor:} Parses Fortran source, builds Fortran ASTs, and supplies parsing and code-generation utilities for the transpilation pipeline;

\item \texttt{Navigator:} Resolves inter-module dependencies through breadth-first traversal, locating declarations, interfaces, external procedures, and module dependencies while prioritizing the shortest path to each search request;

\item \texttt{Extractor:} Analyzes the AST to reconstruct the target procedure's semantic context--including interfaces, dummy arguments, declarations, array dimensions, loops, scopes, call graphs, and inter-procedural dependencies--and produces metadata that underpins subsequent transformations;

\item \texttt{Shaper:} Performs inter-procedural array shape inference, resolving implicit dimensions from declarations, interfaces, and argument propagation across call hierarchies. It reconstructs explicit array specifications--including dimensions inferred via intrinsics such as \texttt{SIZE}--to supply complete shape information for transformations;

\item \texttt{Modifier:} Implements the core \texttt{OPENACC} source-to-source transformation engine, applying deterministic, semantic-preserving rewrites to Fortran constructs--including array operations, implicit shapes, control-flow, procedure interfaces, and loops--directly on the AST;

\item \texttt{Isolator:} Reconstructs the target procedure and its dependency context into a standalone compilation unit for independent execution, validation, debugging, and transpilation.

\end{itemize} 

\begin{table*}[t]
\tabcolsep=0pt%
\TBL{\caption{Comparison between Fortran and Python programming languages.\label{tab:fortran_python_comparison}}}
{\begin{fntable}
\begin{tabular*}{\textwidth}{@{\extracolsep{\fill}}p{2.8cm} p{4.2cm} p{4.2cm}@{}}\toprule
 & \multicolumn{1}{@{}c@{}}{\TCH{\textbf{Fortran}}} & \multicolumn{1}{@{}c@{}}{\TCH{\textbf{Python}}} \\

\textbf{Array Indexing} 
 & n-based: arbitrary lower and upper bounds (e.g., \texttt{a(-7:12)}) 
 & 0-based: arrays start at index 0 \\

\textbf{Memory Layout} 
 & Column-major order 
 & Row-major order; NumPy supports column-major via \texttt{order='F'} \\

\textbf{Array Slicing} 
 & \texttt{a(1:5)} includes indices 1 through 5 (inclusive upper bound) 
 & \texttt{a[0:5]} includes indices 0 through 4 (exclusive upper bound) \\

\textbf{Performance} 
 & Compiled for high-performance computing 
 & Interpreted by default; comparable performance via NumPy/JAX \\

\textbf{Mutability} 
 & All arguments are mutable and updated in place 
 & NumPy arrays are mutable in place; scalars are immutable--modified scalars must be explicitly returned \\

\textbf{Typing System} 
 & Strongly, statically typed with explicit precision control 
 & Dynamically typed; precision must be set explicitly (e.g., \texttt{np.float64}) to match Fortran \\

\textbf{Argument Passing} 
 & Pass-by-reference; procedures modify arguments directly 
 & Pass-by-object-reference; mutable and immutable types must be handled separately to emulate in-place modification \\

\textbf{Execution Model} 
 & Compiled 
 & Interpreted \\

\textbf{Variable Scope} 
 & Implicit scoping rules; modules and \texttt{CONTAINS} define visibility 
 & Explicit scoping via functions, classes, and namespaces \\

\textbf{Array Operations} 
 & Native array operations with elemental functions and array syntax 
 & Native operations via NumPy; Python built-ins do not support vectorized operations \\

\textbf{Data Types} 
 & Native support for complex, character, logical, and user-defined types 
 & Native types include int, float, complex, bool, str; arrays via NumPy \\

\textbf{Parallelism} 
 & Native support via Coarrays, MPI, OpenMP, OpenACC 
 & Via external libraries (e.g., MPI for Python, JAX, CuPy, Numba) \\

\botrule
\end{tabular*}%
\end{fntable}}
\end{table*}

\subsection{Transpilation}\label{subsection:transpilation}

Fortran and Python differ in language semantics, execution model, memory representation, and variable management. As such, semantic equivalence cannot be achieved through direct syntactic translation alone; rather, each language difference is addressed through a corresponding transformation (Table~\ref{tab:fortran_python_comparison}).

The transpilation pipeline reconciles these differences through AST rewriting rules. Control-flow constructs are adapted to zero-based indexing; array allocation and slicing semantics are reconstructed while preserving Fortran memory ordering where needed, with numerical types mapped to NumPy equivalents (e.g., \texttt{np.float64}). Procedure interfaces are rewritten to convert mutable scalar arguments into explicit return values, aligning with Python calling conventions.

\section[NumPy pipeline]{NumPy pipeline} 

The dependency-aware isolation stage produces standalone Fortran procedures with the metadata required for transpilation. Building on these units, the next stage generates executable Python scripts while preserving the separation between global state, procedure logic, and execution flow. The generated program comprises a \texttt{global} class encapsulating transpiled dependencies, global variables, module state, and initialization routines; and a \texttt{main} module orchestrating the execution. Encapsulating shared state within a class--rather than using a module--improves modularity, maintainability, and compatibility with subsequent transformations (Figure~\ref{fig:flowchart_gpt}). The transpilation comprises two passes:

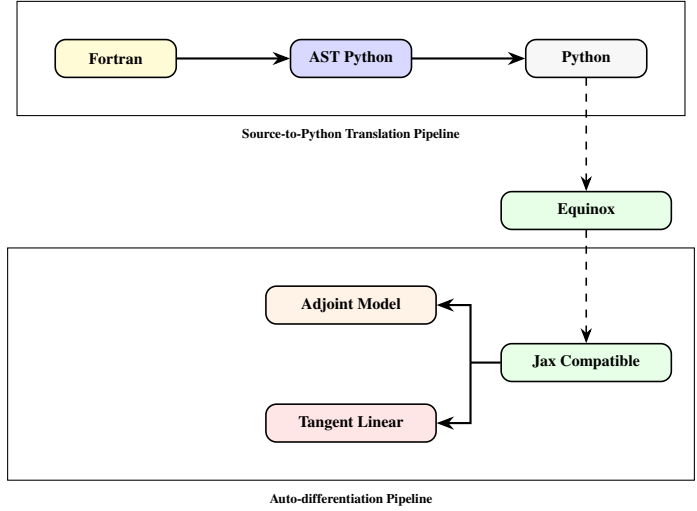
\begin{figure}[t]
        \centering
        \begin{tikzpicture}[
            node distance=3cm and 3cm,
            scale = 0.5,
            transform shape, 
            >=Stealth,
            box/.style={
                draw,
                rectangle,
                minimum height=1cm,
                minimum width=3.2cm,
                align=center,
                font=\large,
                line width=0.6pt,
                rounded corners=4pt
            },
            fortran/.style={box, fill=yellow!20},
            ast/.style={box, fill=blue!15},
            python/.style={box, fill=gray!15},
            equinox/.style={box, fill=green!10, minimum width=4.5cm},
            jax/.style={box, fill=green!10, minimum width=4.5cm},
            adjoint/.style={box, fill=orange!10, minimum width=4.5cm},
            tangent/.style={box, fill=red!10, minimum width=4.5cm},
            arrow/.style={->, line width=0.8pt},
            line/.style={-, line width=0.8pt},
            dottedarrow/.style={->, line width=0.6pt, dashed}
        ]
        
        \node[fortran] (fortran) {\textbf{Fortran}};
        \node[ast, right=of fortran] (ast) {\textbf{AST Python}};
        \node[python, right=of ast] (python) {\textbf{Python}};
        
        \node[
            draw,
            rectangle,
            fit=(fortran)(ast)(python),
            inner xsep=0.5cm,
            inner ysep=0.5cm
        ] (rect1) {};
        
        \node at (rect1.south) [yshift=-0.5cm] {\textbf{Source-to-Python Translation Pipeline}};
        
        \node[equinox, below=of python] (equinox) {\textbf{Equinox}};

        \node[jax, below=of equinox] (jax) {\textbf{Jax Compatible}};
        \node[adjoint, below=5.5cm of ast] (adjoint) {\textbf{Adjoint Model}};
        \node[tangent, below=2.1cm of adjoint] (tangent) {\textbf{Tangent Linear}};

        \node[
            draw,
            rectangle,
            inner ysep=0.5cm,
            fit=
                (rect1.west |- jax)
                (rect1.east |- jax)
                (jax.north)
                (adjoint.north)
                (tangent.south)
        ] (rect2) {};

        \node at (rect2.south) [yshift=-0.5cm] {\textbf{Auto-differentiation Pipeline}};
        
        \draw[arrow] (fortran) -- (ast);
        \draw[arrow] (ast) -- (python);

        \draw[dottedarrow] (python) -- (equinox);
        \draw[dottedarrow] (equinox) -- (jax);

        \coordinate (merge) at ($(jax.west)+(-0.8,0)$);

        \draw[line] (jax.west) -- (merge);
        \draw[arrow] (merge) |- (adjoint.east);
        \draw[arrow] (merge) |- (tangent.east);
        
        \end{tikzpicture}
        \caption{Workflow of the transpiler}
        \label{fig:flowchart_gpt}
    \end{figure}

\begin{itemize} 
    \item \texttt{Transformer}: Assembles the overall Python program by constructing the \texttt{global} class, and \texttt{main} module, generating initialization logic, integrating translated procedures, resolving dependencies, and producing a Python AST.
    
    \item \texttt{F2NP}: Performs the semantic lowering of isolated Fortran procedures into equivalent NumPy implementations. This stage rewrites language-specific constructs--including loop bounds, array indexing, intrinsic procedures, control-flow statements, and procedure interfaces--while preserving the execution logic of the original implementation.
\end{itemize}

Subsequent \texttt{NodeTransformer}-based corrective passes perform additional semantic corrections--including indexing semantics, attribute adjustments, etc--to produce executable Python code. Finally, standardized templates assemble the transformed AST into executable Python scripts.

\begin{figure}[t]
\centering
\begin{tikzpicture}[
   scale = 0.5,
   transform shape, 
    >=Stealth,
    node distance=1.3cm and 2.2cm,
    every node/.style={font=\small},
    stage/.style={
        draw=black!70,
        rectangle,
        rounded corners=3pt,
        minimum width=6.2cm,
        minimum height=1cm,
        align=center,
        line width=0.8pt,
        fill=teal!12,
        text width=6cm
    },
    startend/.style={
        stage,
        fill=teal!35,
        line width=1pt
    },
    decision/.style={
        draw=black!70,
        diamond,
        aspect=2.4,
        align=center,
        line width=0.8pt,
        fill=orange!20,
        inner sep=2pt,
        text width=2.6cm
    },
    branch/.style={
        stage,
        minimum width=4.6cm,
        text width=4.2cm,
        fill=blue!10
    },
    arrow/.style={->, line width=0.9pt, black!70},
    lbl/.style={font=\scriptsize\itshape, black!60}
]

\node[startend] (input) {Fortran AST Node\\[1pt] {\scriptsize (Subroutine / Function)}};
\node[stage, below=of input] (dispatch)
    {Send via \texttt{recursive\_ast}\\[1pt]
     {\scriptsize\texttt{visit\_Subroutine} / \texttt{visit\_Function}}};
\node[stage, below=of dispatch] (signature)
    {Extract Signature\\[1pt] {\scriptsize Name, Arguments, Types}};
\node[stage, below=of signature] (args)
    {Argument Normalization\\[1pt] {\scriptsize Intent handling, defaults}};
\node[stage, below=of args] (body)
    {Recursive Body Transformation\\[1pt] {\scriptsize Statements \& Expressions}};
\node[decision, below=1.5cm of body] (returncheck)
    {Function\\ or\\ Subroutine?};

\node[branch, below left=1.6cm and 1 cm of returncheck] (func)
    {Generate Return\\Statement};
\node[branch, below right=1.6cm and 1 cm of returncheck] (sub)
    {Generate Procedure\\Body Only};

\node[stage, below=3.3cm of returncheck] (emit)
    {Generate Python \texttt{ast.FunctionDef}};
\node[startend, below=of emit] (store)
    {Append to Module Body};

\draw[arrow] (input) -- (dispatch);
\draw[arrow] (dispatch) -- (signature);
\draw[arrow] (signature) -- (args);
\draw[arrow] (args) -- (body);
\draw[arrow] (body) -- (returncheck);

\draw[arrow] (returncheck.west) -| node[lbl, pos=0.25, above] {Function} (func.north);
\draw[arrow] (returncheck.east) -| node[lbl, pos=0.25, above] {Subroutine} (sub.north);

\draw[arrow] (func.south) |- (emit.west);
\draw[arrow] (sub.south) |- (emit.east);

\draw[arrow] (emit) -- (store);

\end{tikzpicture}
\caption{Workflow of the \texttt{F2NP} transpiler}
\label{fig:f2np_workflow}
\end{figure}
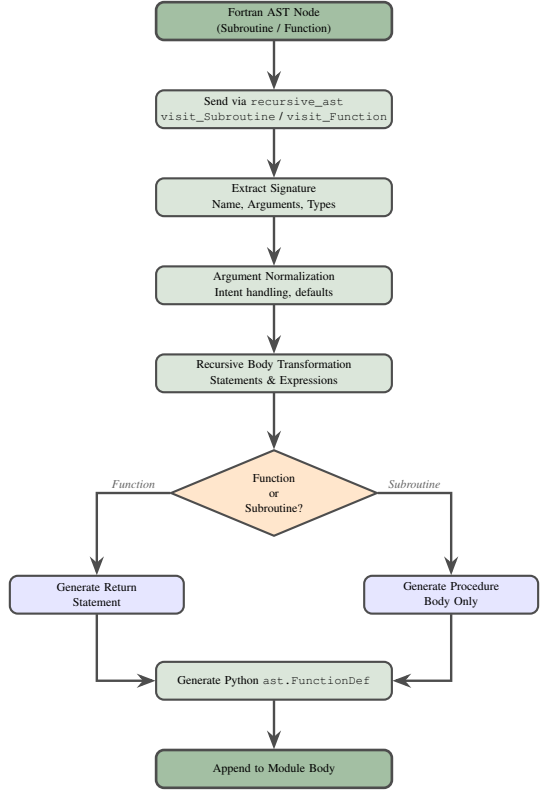

\subsection[F2NP transpiler]{F2NP transpiler} \label{sec:f2np}
The \texttt{F2NP} stage lowers isolated Fortran procedures to equivalent NumPy implementations. It constructs a Python AST that preserves the original control-flow, procedure interfaces, and computational semantics. Figure~\ref{fig:f2np_workflow} illustrates this process, from the Fortran AST to the generated \texttt{ast.FunctionDef}. The transpiler uses Fortran AST, where each node includes parent and child relationships--enabling recursive traversal and coherent reconstruction of the equivalent Python AST.

A principal challenge arises from the differing control-flow representations in Fortran and Python. Fortran explicitly terminates structures with \texttt{END} statements, whereas Python denotes block structure via indentation. The transpiler exploits the explicit termination statements in the Fortran AST to rebuild the Python hierarchy using two last-in-first-out (LIFO) stacks:

\begin{itemize} 
    \item A module stack that constructs the overall function body;
    \item A control stack that tracks the current nesting context of conditionals, loops, and other control-flow constructs.
\end{itemize}

When a new control-flow construct is encountered, its corresponding Python AST node is created, attached to the current parent's body, and pushed onto the control stack. Subsequent statements are appended to the top-of-stack node, allowing nested structures to be reconstructed incrementally during recursive traversal. Since the stack stores references to live AST nodes, modifications propagate immediately throughout the tree, obviating additional reconstruction passes.

Conditional constructs require special handling, as some Fortran control-flow statements lack explicit terminators. An \texttt{ELSEIF} branch implicitly closes the preceding \texttt{IF} branch without an \texttt{END} statement; \texttt{ELSEWHERE} behaves similarly for \texttt{WHERE} constructs. To recover these implicit boundaries, the stack-based traversal is augmented with independent nesting counters for \texttt{DO}, \texttt{IF}, \texttt{ELSEIF}, \texttt{WHERE}, \texttt{ELSEWHERE}, and \texttt{CASE} constructs. These counters track the active nesting level of each type, enabling correct hierarchical reconstruction regardless of nesting complexity (Figure~\ref{fig:if_control}, Appendix~\ref{appendixA}).

With the control-flow hierarchy reconstructed, the \texttt{F2NP} transpiler applies semantic rewrite rules to resolve language-specific differences between Fortran and Python. These transformations operate locally on the AST and do not require the entire program context:

\begin{itemize} 
    \item Loop-bound: Fortran loop bounds are adjusted for n-based to zero-based indexing in Python \texttt{range} expressions.
    
    \item Array-dimension: Fortran array bounds are transformed to NumPy dimensions via $\text{size} = \text{upperbound} - \text{lowerbound} + 1$.
    
    \item Intrinsic-function: Fortran intrinsics (\texttt{SUM}, \texttt{MIN}, \texttt{MAX}, etc.) are mapped to NumPy equivalents, with \texttt{DIM} translated to \texttt{axis} regardless of explicitness.
\end{itemize}


\subsection[Transformer]{Transformer}\label{sec:transformer} 

The \texttt{F2NP} stage lowers isolated Fortran procedures to NumPy, while the \texttt{Transformer} stage assembles these components into complete Python scripts. As the orchestration layer, it constructs the top-level AST, integrates transpiled procedures, generates initialization logic, and produces the final program structure for validation and evaluation (Figure~\ref{fig:full_pipeline}).

Using structural metadata from isolation, the \texttt{Transformer} constructs the NumPy code into two components: a \texttt{global} class encapsulating shared variables, module state, initialization, and translated dependency procedures; and a \texttt{main} module handling execution, input, and validation.

\subsubsection[Global class]{Global class}\label{sec:global_class} 

Construction of the \texttt{global} class proceeds in two stages: structural assembly and semantic assembly. First, global variables required by the isolated procedure are assembled--declarations, initial values, array dimensions, and input layouts are extracted from Fortran and translated to Python. Next, the transpiled procedures from \texttt{F2NP} are assembled. Dependencies are resolved via depth-first traversal of the call graph, ensuring lower-level routines are incorporated before their callers--preserving execution semantics while producing a modular, object-oriented representation. 

\subsubsection[Main module]{Main module}\label{sec:main_module} 
The \texttt{main} module serves as the entry point for each isolated procedure, orchestrating environment initialization, procedure invocation, and numerical validation against the original Fortran implementation. Its construction proceeds in three stages: (i) analyzing the \texttt{global} class interface for correct access to procedures and state; (ii) initializing local variables and input data using reference datasets from isolation; and (iii) automatically generating validation routines to compare Python outputs against Fortran reference values. The \texttt{main} module also incorporates runtime measurements, enabling performance comparisons between the generated Python implementation and its Fortran counterpart.

\subsection[Corrective passes]{Corrective passes} \label{sec:normalization}
Although \texttt{F2NP} preserves structural semantics, certain language-specific corrections--array lower bounds, class attribute mappings, and object-composition relationships--require semantic information that only becomes available after program assembly. FGPT therefore applies two successive corrective passes to the generated Python AST: \texttt{AdjustIndices}, which restores Fortran indexing semantics; and \texttt{ReplaceGlobals}, which reconstructs object-oriented attribute and method references.

\subsubsection{Array index correction} 
The first corrective pass, \texttt{AdjustIndices} (Figure~\ref{fig:adjust_indices_full}), restores Fortran indexing semantics within Python's zero-based framework. Driven by array metadata from isolation, each array is corrected according to its declared bounds. Arrays with the default lower bound of one have index expressions shifted by one position; those with arbitrary bounds (e.g., \texttt{0:n} or \texttt{-3:n}) require an explicit offset of (1 - lower bound), applied only to loop variables and identified index expressions.

Beyond explicit indexing, functions like \texttt{argmin} and \texttt{argmax}--which return zero-based indices in NumPy but one-based in Fortran--are adjusted by restoring the appropriate lower bound of the referenced array. For multidimensional arrays, the correction is derived from the \texttt{axis} argument to ensure dimensional correctness. To maintain consistency, variables with corrected indices are tracked; subsequent uses in assignments, loop bounds, conditionals, and comparisons reuse the existing correction rather than reapplying the offset.

\begin{figure}[t]
\centering
\begin{tikzpicture}[
scale=0.5,
transform shape,
>=Latex,
node distance=8mm and 16mm,
method/.style={
    draw,
    rounded corners,
    fill=blue!15,
    minimum width=3.8cm,
    minimum height=8mm,
    align=center,
    font=\small
},
helper/.style={
    draw,
    rounded corners,
    fill=green!10,
    minimum width=3.8cm,
    minimum height=8mm,
    align=center,
    font=\small
},
data/.style={
    draw,
    rounded corners,
    fill=yellow!18,
    minimum width=4.2cm,
    minimum height=10mm,
    align=center,
    font=\small
},
stage/.style={draw, rounded corners, align=center, font=\scriptsize, text width=2.5cm, minimum height=1.2cm},
group/.style={
        draw,
        dashed,
        rounded corners,
        inner sep=6pt
    },
groupbox1/.style={draw, rectangle, rounded corners, inner sep=10pt},
]

\node[data] (meta)
{
\textbf{Shared Metadata}\\[1mm]
\texttt{array\_info}\\
\texttt{CONV\_VARS}\\
\texttt{adjusted\_vars}
};

\node[method, below right=5mm and 15mm of meta] (sub)
{visit\_Subscript};
\node[method, below=of sub] (assign)
{visit\_Assign};
\node[method, below=of assign] (forloop)
{visit\_For};
\node[method, below=of forloop] (iff)
{visit\_If};
\node[method, below=of iff] (call)
{visit\_Call};

\node[helper, right=38mm of assign] (adj)
{\_adjust\_index};
\node[helper, above=of adj] (offset)
{\_apply\_offset\_if\_convvar};
\node[helper, below=of adj] (rhs)
{\_adjust\_assignment\_rhs};
\node[helper, below=of rhs] (compare)
{\_handle\_compare};

\node[helper, below=of compare] (binop)
{\_handle\_binop};

\foreach \n in {sub,assign,forloop,iff,call}
{
    \draw[dashed,->] (meta.south) |- (\n.west);
}

\draw[->] (sub.east) -- ++(8mm,0) |- (adj.west);
\draw[->] (sub.east) -- ++(8mm,0) |- (offset.west);
\draw[->] (assign.east) -- ++(8mm,0) |- (adj.west);
\draw[->] (assign.east) -- ++(8mm,0) |- (rhs.west);
\draw[->] (forloop.east) -- ++(8mm,0) |- (adj.west);
\draw[->] (iff.east) -- ++(8mm,0) |- (compare.west);
\draw[->] (call.east) -- ++(8mm,0) |- (offset.west);


\draw[->] (adj.east) -- ++(8mm, 0) |- (binop.east);

\node[group,
fit=(sub)(assign)(forloop)(iff)(call),
label=above:{\textbf{AST Visitor Methods}}
](group1){};

\node[group,
fit=(offset)(adj)(rhs)(compare)(binop),
label=above:{\textbf{Index Transformation Helpers}}
](group2){};

\node[groupbox1, 
fit=(group2)(group1)(meta),
label=below:{\textbf{AdjustIndices}}
](group3){};

\node[stage, fill=gray!15, left=2.5cm of group3] (input) {Python/NumPy\\AST input};

\node[stage, fill=gray!15, right=2.5cm of group3] (output)
{\textbf{Adjusted Python AST}};

\draw[->] (input) -- (group3);
\draw[->] (group3) -- (output);

\end{tikzpicture}
\caption{\texttt{AdjustIndices} pass}
\label{fig:adjust_indices_full}
\end{figure}

\subsubsection{Attribute resolution} 
The second corrective pass, \texttt{ReplaceGlobals} (Figure~\ref{fig:replace_globals}), reconstructs the object-oriented structure introduced during program assembly. Because variables and procedure calls are translated before the class hierarchy is established, references to global state remain unattributed until the complete program structure is available. Once assembly is complete, structural metadata from isolation resolves these references: variables belonging to the \texttt{global} class are rewritten as \texttt{self.attribute}, and standalone function calls are converted to attributed method calls. This restores the shared state–procedure relationships established during reconstruction while conforming to Python's object-oriented model.

Attribute resolution applies uniformly across all expression contexts--assignments, loop bounds, conditionals, Boolean operations, comparisons, and formatted strings--ensuring consistent representation of shared state throughout the generated program.

\begin{figure}[t]
\centering
\begin{tikzpicture}[
scale=0.5,
transform shape,
>=Latex,
node distance=8mm and 18mm,
method/.style={
    draw,
    rounded corners,
    fill=blue!15,
    minimum width=3.8cm,
    minimum height=8mm,
    align=center,
    font=\small
},
helper/.style={
    draw,
    rounded corners,
    fill=green!12,
    minimum width=3.8cm,
    minimum height=8mm,
    align=center,
    font=\small
},
data/.style={
    draw,
    rounded corners,
    fill=yellow!18,
    minimum width=4.2cm,
    minimum height=10mm,
    align=center,
    font=\small
},
stage/.style={draw, rounded corners, align=center, font=\scriptsize, text width=2.5cm, minimum height=1.2cm},
group/.style={
        draw,
        dashed,
        rounded corners,
        inner sep=6pt
    },
groupbox1/.style={draw, rectangle, rounded corners, inner sep=6pt},
]

\node[data] (meta)
{
\textbf{Metadata}\\
\texttt{cls\_info}\\
\texttt{\_local\_scope}
};

\node[method, below right=10mm and 10mm of meta] (name)
{visit\_Name};

\node[method, below=of name] (attr)
{visit\_Attribute};

\node[method, below=of attr] (assign)
{visit\_Assign};

\node[method, below=of assign] (call)
{visit\_Call};

\node[method, below=of call] (forloop)
{visit\_For};

\node[method, below=of forloop] (iff)
{visit\_If};

\node[method, below=of iff] (binop)
{visit\_BinOp};

\node[helper, right=42mm of assign] (resolver)
{get\_attr\_node};

\node[helper, below=of resolver] (compare)
{\_replace\_compare};

\foreach \m in {name,attr,assign,call,forloop,iff,binop}
{
    \draw[dashed,->] (meta.south) |- (\m.west);
}

\foreach \m in {name,attr,assign,call,forloop,iff,binop}
{
    \draw[->] (\m.east) -- ++(8mm,0) |- (resolver.west);
}

\draw[->]
(iff.east)
-- ++(8mm,0)
|- (compare.west);

\draw[->]
(compare.north)
-- (resolver.south);

\node[group,
fit=(name)(attr)(assign)(call)(forloop)(iff)(binop),
label=above:{\textbf{AST Visitor Methods}}
](group1){};

\node[group,
fit=(resolver)(compare),
label=above:{\textbf{Name Resolution Helpers}}
](group2){};

\node[groupbox1, 
fit=(group2)(group1)(meta),
label=below:{\textbf{ReplaceGlobals}}
](group3){};

\node[stage, fill=gray!15, left=2.5cm of group3] (input) {Python/NumPy\\AST input};

\node[stage, fill=gray!15, right=18mm of group2] (output)
{\textbf{Rewritten Python AST}};

\draw[->] (input) -- (group3);
\draw[->] (group3) -- (output);

\end{tikzpicture}
\caption{\texttt{ReplaceGlobals} pass}
\label{fig:replace_globals}
\end{figure}

\section{Numerical verification} \label{sec:validation}
We verify that FGPT preserves the numerical precision of the original Fortran code when generating NumPy and JAX implementations by comparing outputs against the original Fortran. Since each procedure is isolated prior to transformation, reference inputs and outputs are generated for every translation unit, enabling translation errors to be localized to individual passes rather than only after full program generation. The same validation procedure is applied to both NumPy and JAX implementations. Numerical consistency is quantified using the maximum absolute error,
\[
\max\left(\left|Y-Y_{\text{F}}\right|\right),
\]
where $Y$ denotes the output of the transpiled implementation (NumPy or JAX) and $Y_{\text{F}}$ is the corresponding Fortran output. In practice, this criterion is evaluated using \texttt{numpy.isclose} (scalars and Booleans) and \texttt{numpy.allclose} (arrays), both applying \[ |a-b| \leq \text{atol} + \text{rtol}\cdot|b|, \] where \(a\) is the either NumPy or JAX implementation and \(b\) the Fortran reference. Default tolerances are \(\text{rtol}=1\times10^{-5}\) and \(\text{atol}=1\times10^{-8}\). All benchmarks were performed on CPU unless otherwise stated.

\subsection{Benchmark}
To evaluate the proposed pipeline, we consider the \texttt{hydrology} module from the IPSL land-surface model \citep{boucher_2020}, comprising around 15,000 lines of Fortran code with the following procedures:

\begin{verbatim}
    hydrol_main/
    |-- hydrol_soil
    |-- hydrol_vegupd
    |-- hydrol_alma
    |-- hydrol_canop
    |-- explicitsnow_main
    |-- hydrol_hydraulic_arch_tuzet_calc
\end{verbatim}

Several of these procedures call additional child procedures, which are recursively isolated and transpiled as part of the dependency analysis. Consequently, the entire \texttt{hydrology} workflow is translated into both NumPy and JAX. These procedures include multidimensional arrays, nested procedure hierarchies, conditional execution, iterative loops, intrinsic functions, and extensive module dependencies, providing a representative evaluation on a real-world scientific application.

\begin{figure}[t]
    \centering
    \begin{subfigure}[b]{0.45\textwidth}
        \centering
        \includegraphics[width=\textwidth]{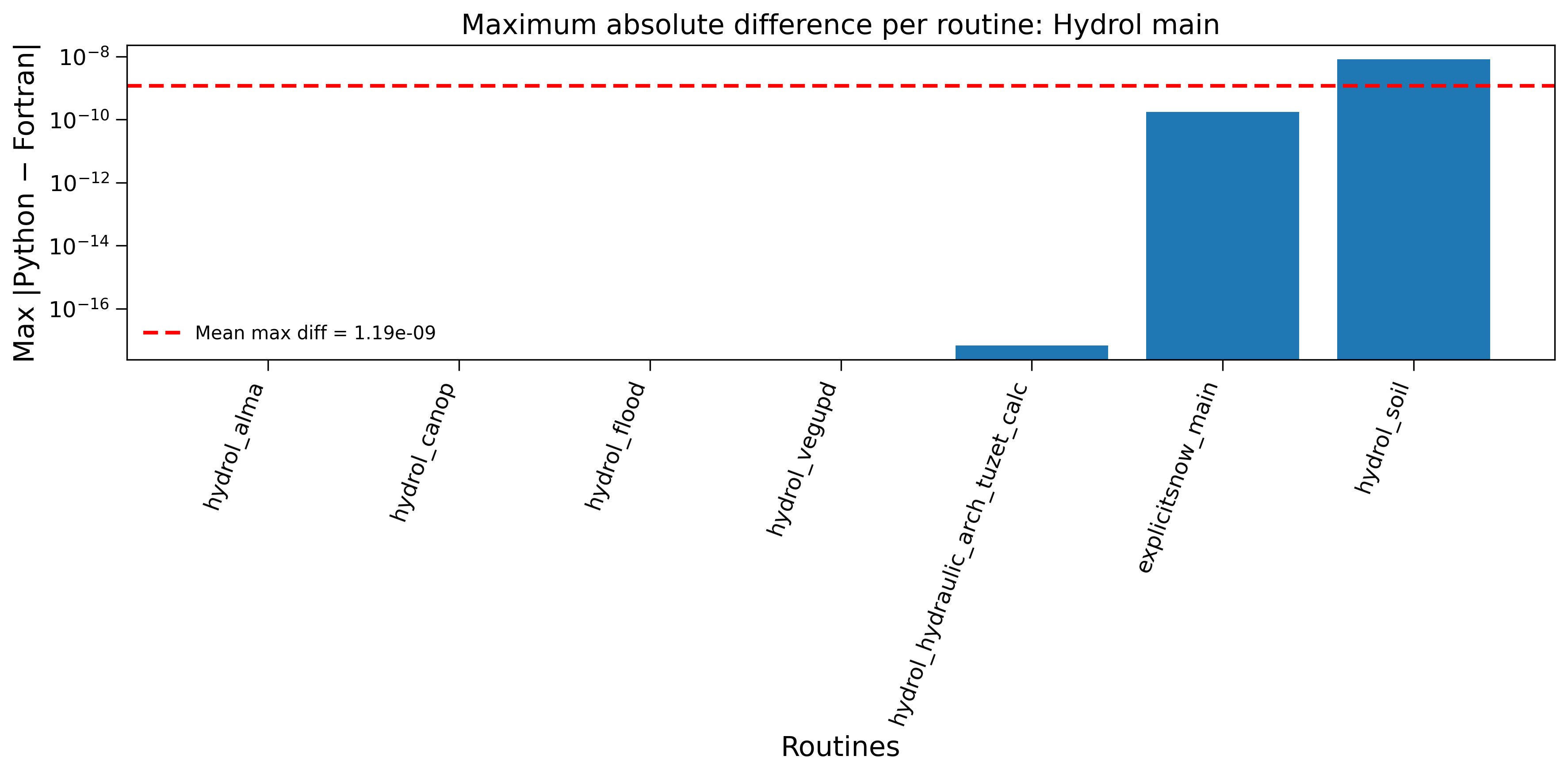}
        \caption{\texttt{hydrol\_main}.}
        \label{fig:absdiff_hydrol_main}
    \end{subfigure}
    \hfill
    \begin{subfigure}[b]{0.45\textwidth}
        \centering
        \includegraphics[width=\textwidth]{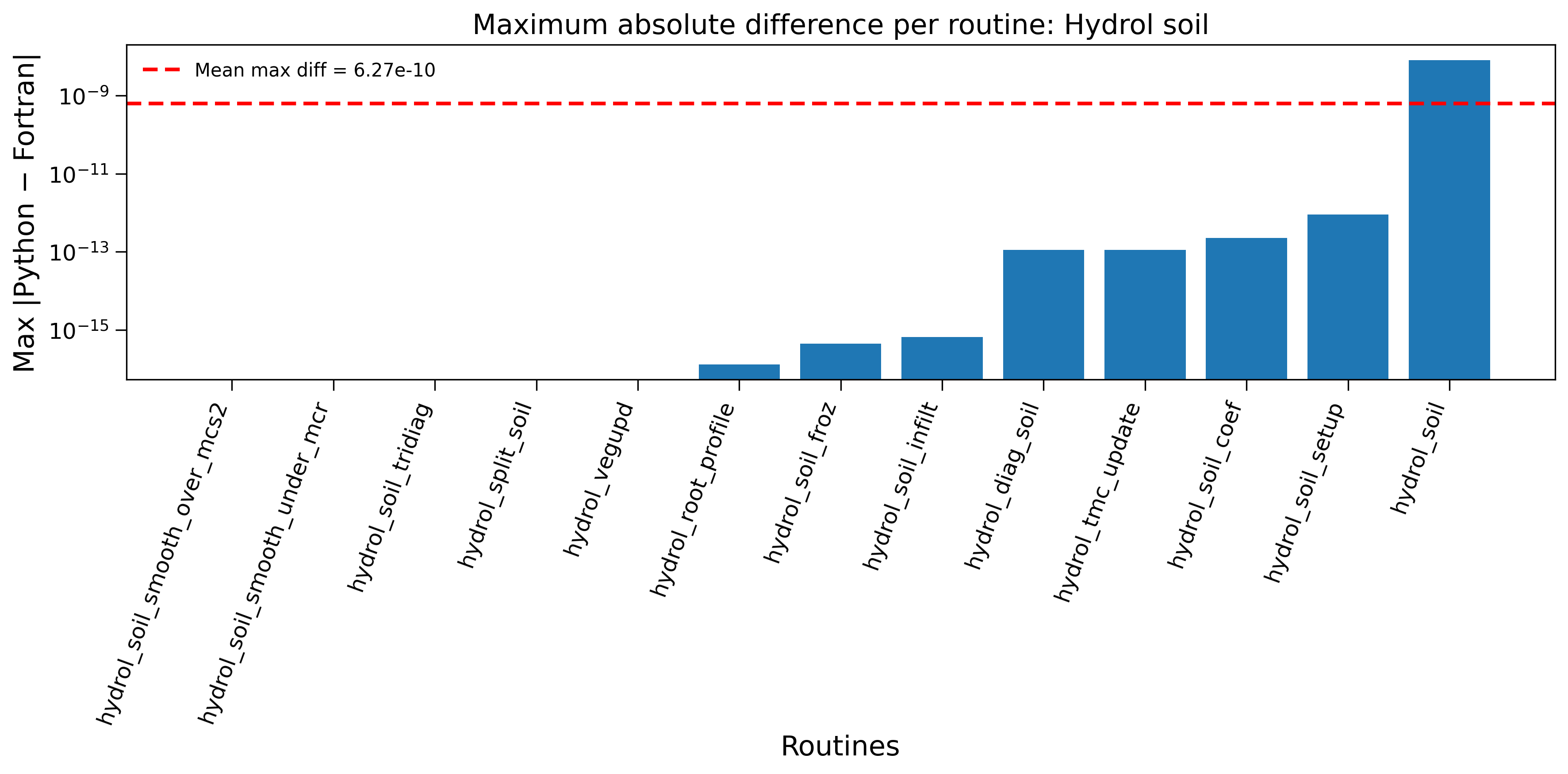}
        \caption{\texttt{hydrol\_soil}.}
        \label{fig:absdiff_hydrol_soil}
    \end{subfigure}
    \vskip\baselineskip
    \begin{subfigure}[b]{0.45\textwidth}
        \centering
        \includegraphics[width=\textwidth]{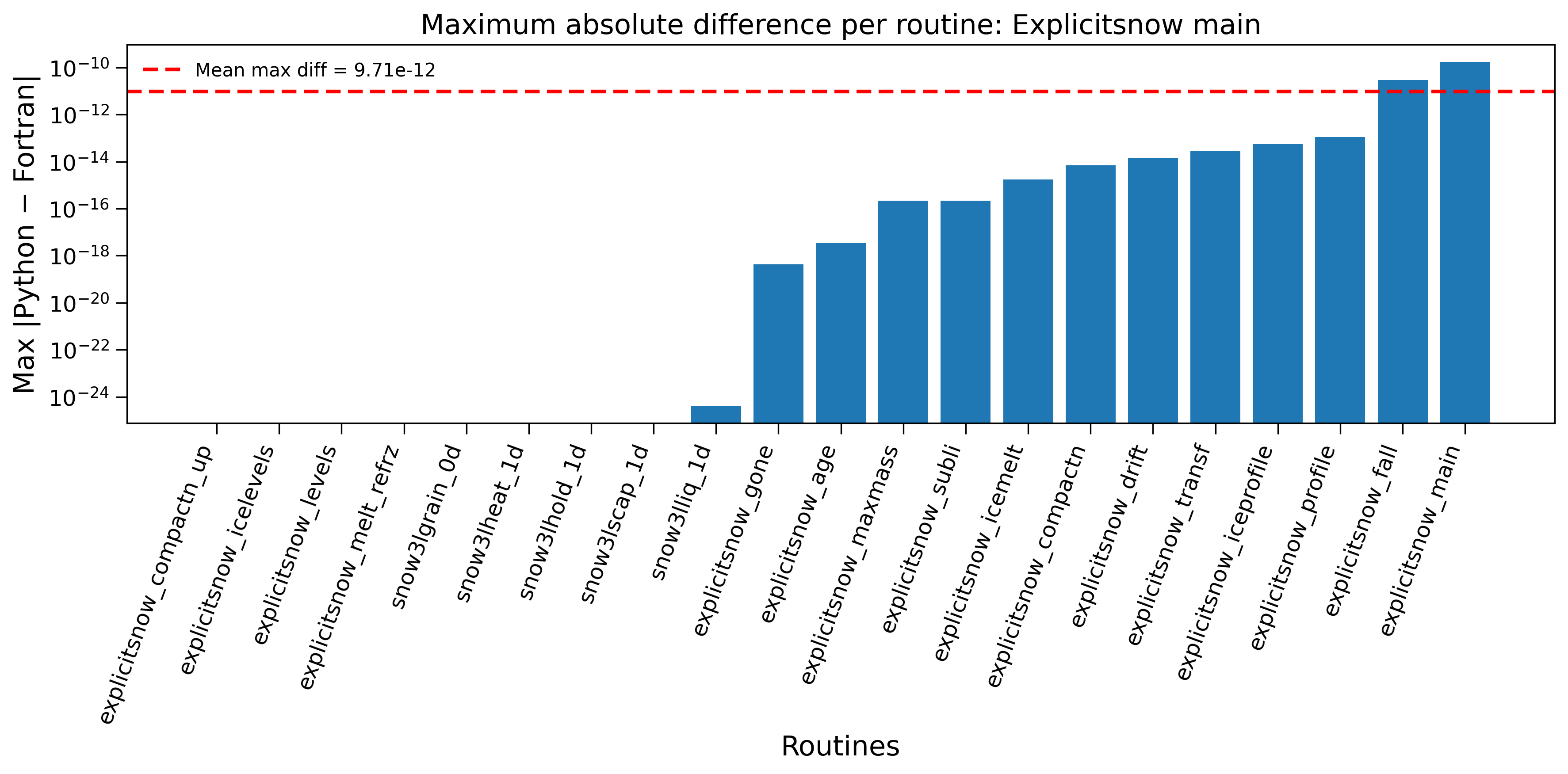}
        \caption{\texttt{explicitsnow\_main}.}
        \label{fig:absdiff_explicitsnow_main}
    \end{subfigure}
    \hfill
    \begin{subfigure}[b]{0.45\textwidth}
        \centering
        \includegraphics[width=\textwidth]{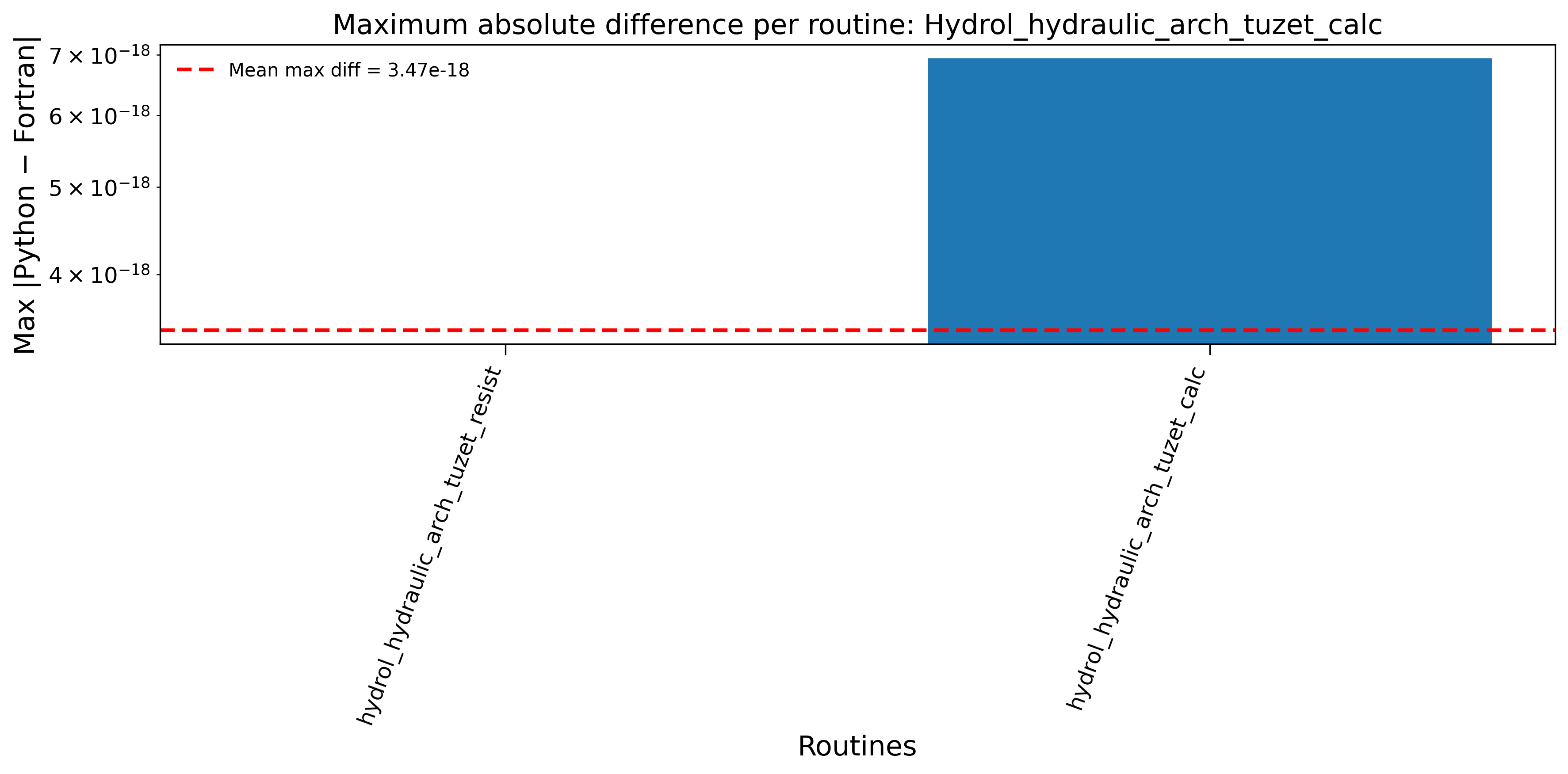}
        \caption{ \texttt{hydrol\_hydraulic\_arch\_tuzet\_calc}.}
        \label{fig:absdiff_hydrol_hydraulic_calc}
    \end{subfigure}

    \caption{Maximum absolute difference between Numpy and Fortran outputs}
    \label{fig:max_abs}
\end{figure}

For each isolated procedure, numerical consistency is quantified by reporting the maximum absolute difference between the original Fortran outputs and the corresponding transpiled implementation. Complete results are presented in Figures~\ref{fig:absdiff_hydrol_main}--\ref{fig:absdiff_hydrol_hydraulic_calc}.

Across all evaluated procedures, the maximum absolute difference is typically on the order of \(10^{-9}\). Differences of this magnitude are expected between the two implementations and arise primarily from two sources. First, floating-point arithmetic is not associative; mathematically equivalent implementations that perform operations in different orders may accumulate rounding errors differently. Second, although both implementations conform to the floating-point standard, they employ different execution strategies. Optimizing Fortran compilers perform instruction scheduling, vectorization, and architecture-specific optimizations, whereas NumPy delegates many operations to optimized BLAS and LAPACK libraries. These implementation differences can produce variations in floating-point rounding while remaining numerically equivalent.

\begin{figure}[t]
    \centering
    \begin{subfigure}[b]{0.45\textwidth}
        \centering
        \includegraphics[width=\textwidth]{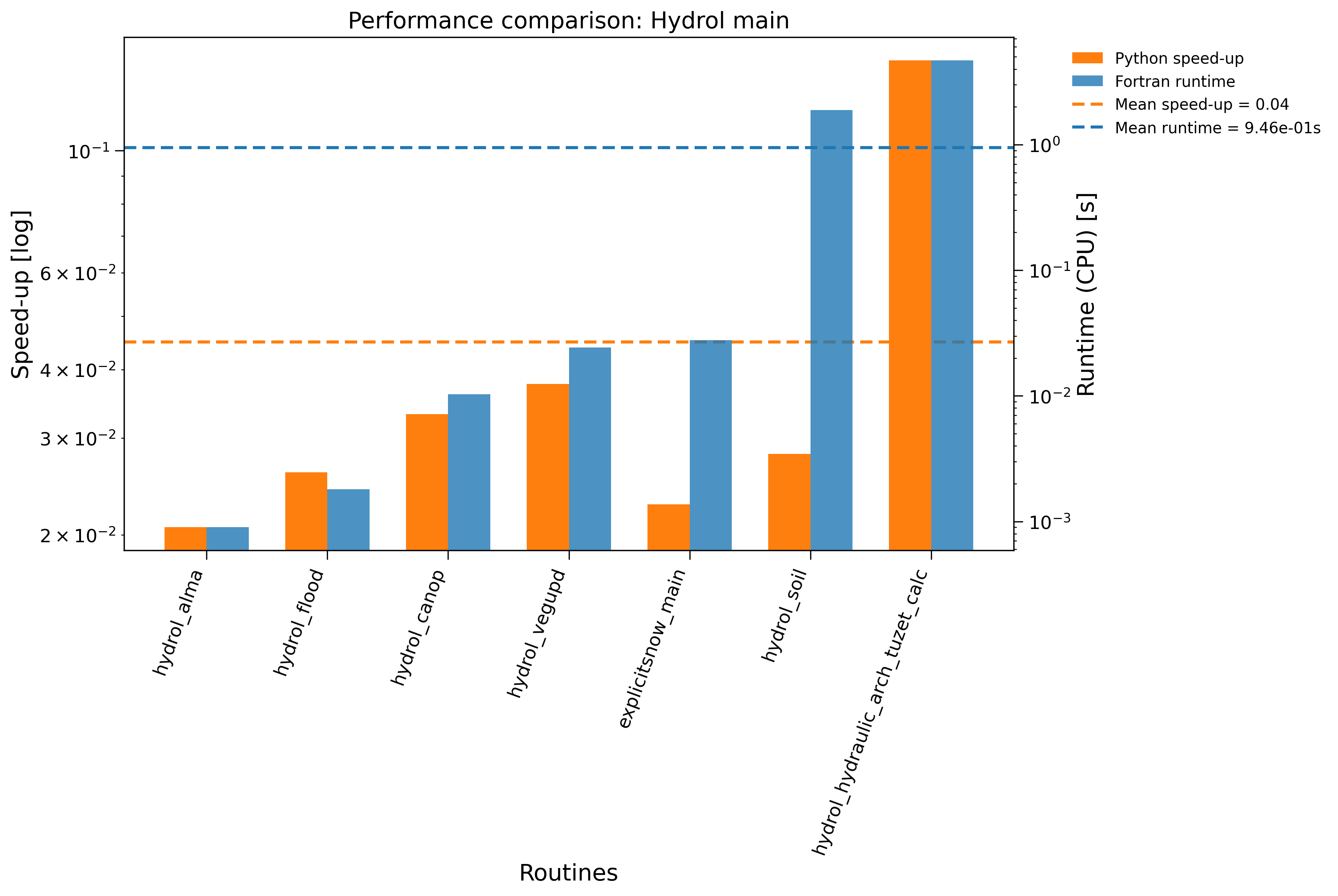}
        \caption{\texttt{hydrol\_main}.}
        \label{fig:perf_hydrol_main}
    \end{subfigure}
    \hfill
    \begin{subfigure}[b]{0.45\textwidth}
        \centering
        \includegraphics[width=\textwidth]{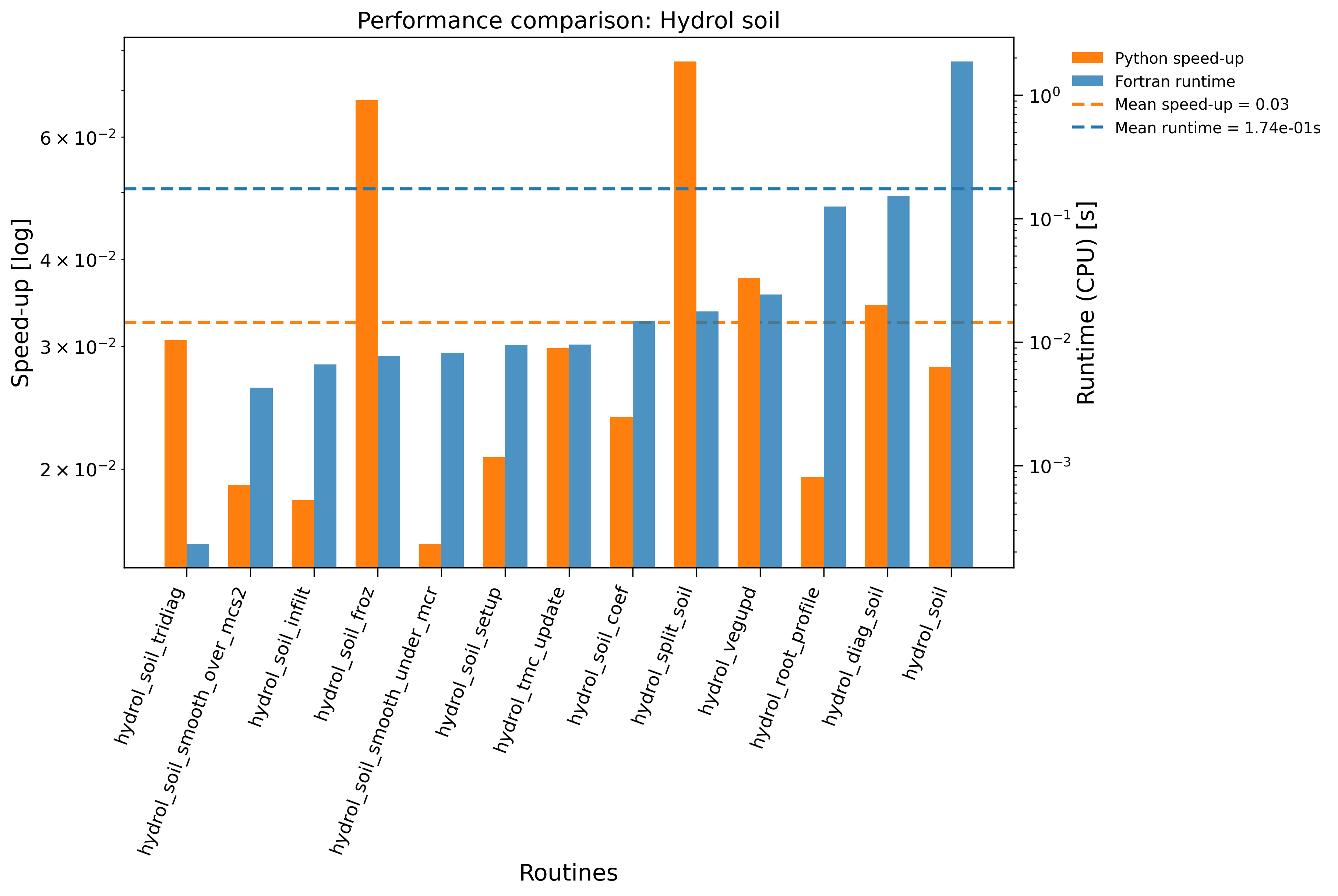}
        \caption{\texttt{hydrol\_soil}.}
        \label{fig:perf_hydrol_soil}
    \end{subfigure}
    \vskip\baselineskip
    \begin{subfigure}[b]{0.45\textwidth}
        \centering
        \includegraphics[width=\textwidth]{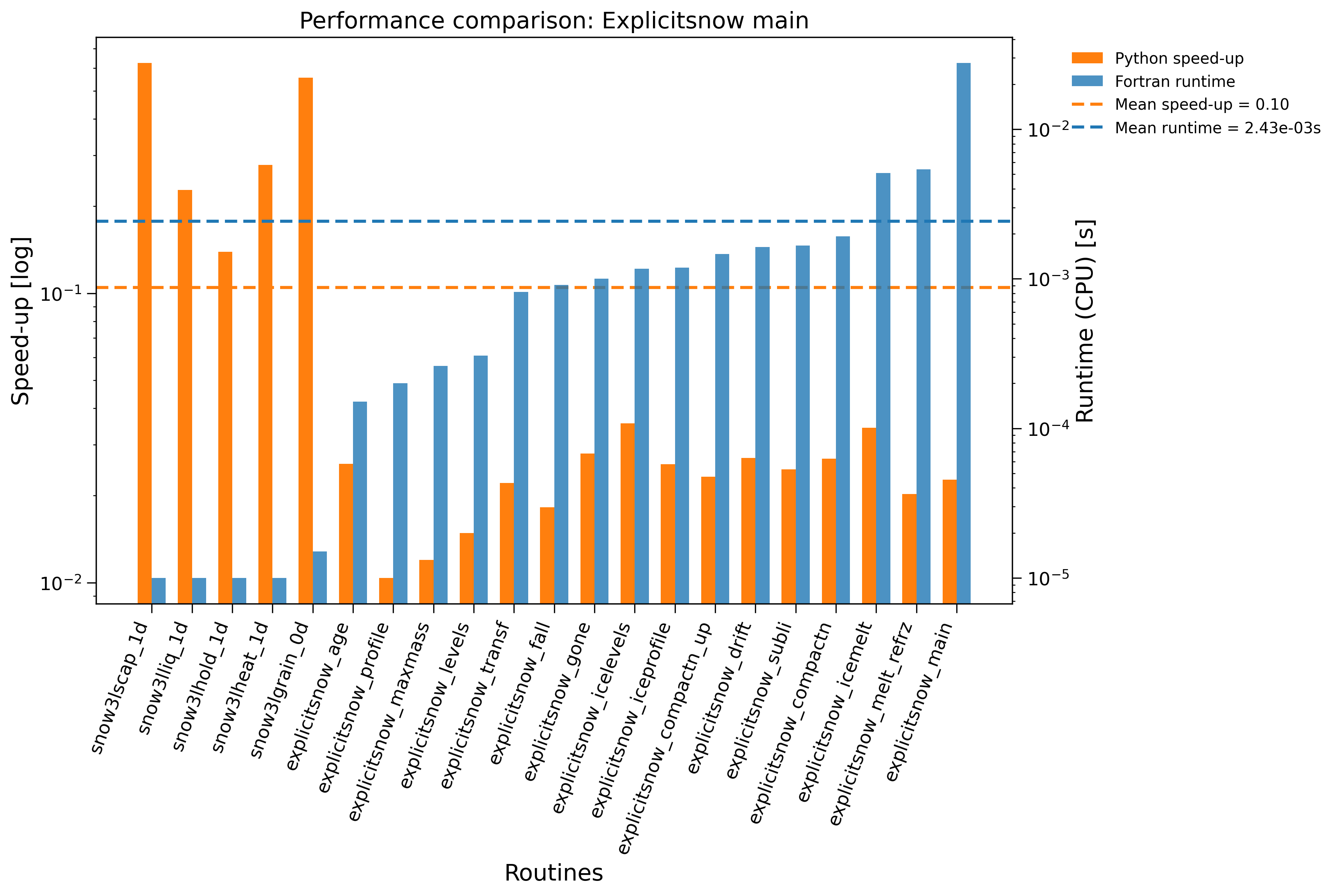}
        \caption{\texttt{explicitsnow\_main}.}
        \label{fig:perf_explicitsnow_main}
    \end{subfigure}
    \hfill
    \begin{subfigure}[b]{0.45\textwidth}
        \centering
        \includegraphics[width=\textwidth]{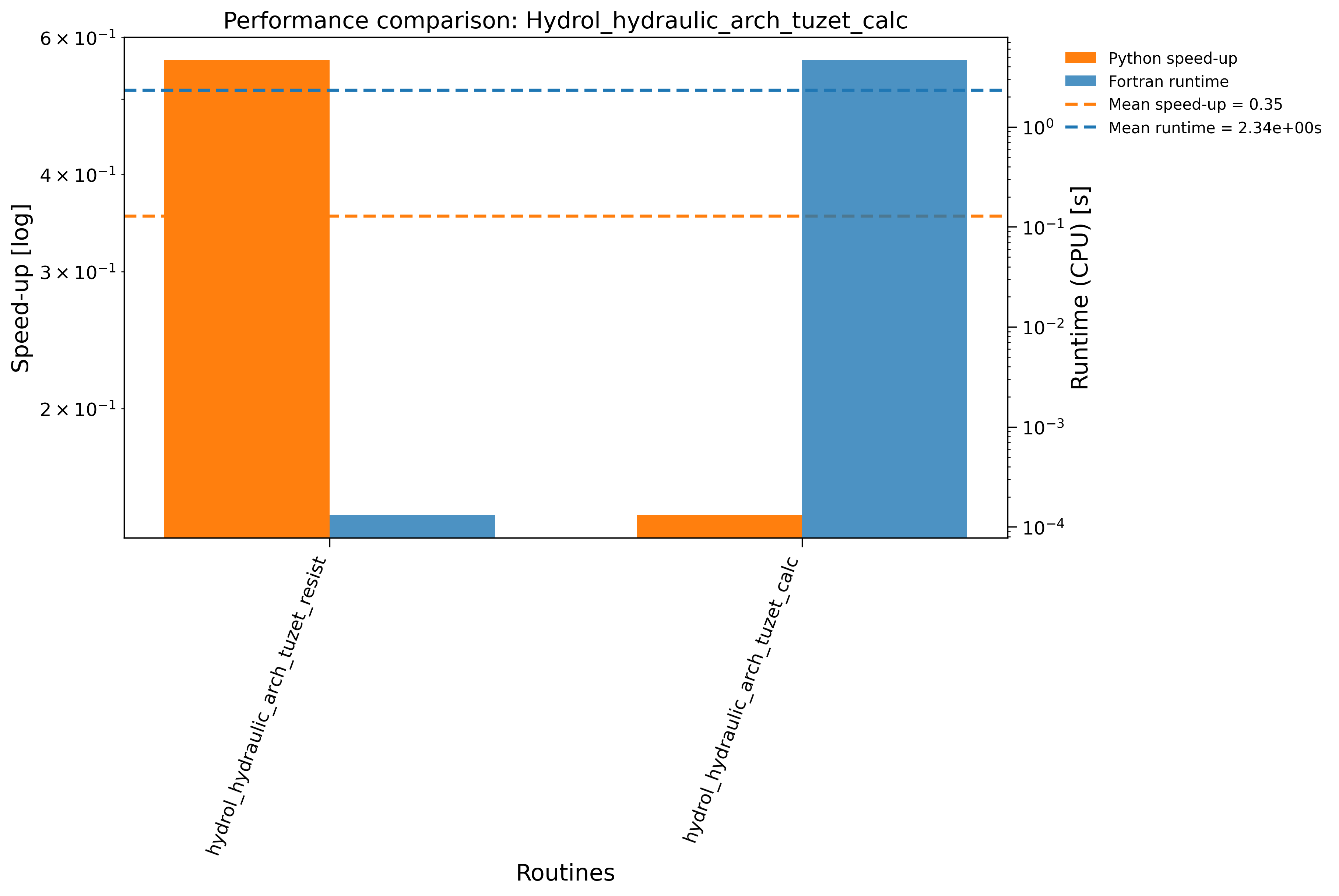}
        \caption{\texttt{hydrol\_hydraulic\_arch\_tuzet\_calc}.}
        \label{fig:perf_hydrol_hydraulic_calc}
    \end{subfigure}

    \caption{Runtime and speedup comparison between Fortran and Numpy implementations}
    \label{fig:speed_up}
\end{figure}

The performance is evaluated by comparing wall-clock execution time between the Fortran and NumPy implementations. Speedup is computed as
\[
\text{Speedup} = \frac{T_{\text{F}}}{T},
\]
where \(T_{\text{F}}\) and \(T\) denote the measured execution times of the Fortran and NumPy/Jax implementations, respectively.  Figure~\ref{fig:perf_hydrol_main} shows execution time and speedup for \texttt{hydrology}; child-procedure results are shown in Figures~\ref{fig:perf_hydrol_soil}--\ref{fig:perf_hydrol_hydraulic_calc}.

The NumPy implementation is order or magnitude slower than the original Fortran code, which is expected: Fortran is compiled by a compiler into efficient machine code, while the NumPy implementation executes through the Python interpreter, incurring additional overhead from interpretation, function calls, and object management. The performance gap is not uniform across procedures, however; where a procedure's core computation maps cleanly onto NumPy's vectorized primitives, the NumPy version approaches--and in some cases matches--Fortran's performance. Conversely, procedures dominated by deeply nested loops or iterative updates favor Fortran's compiled execution model.

\section{Towards differentiable programming with JAX}
\label{sec:jax}

JAX \citep{jax2018github} performs AD by tracing Python function execution to build a JAXPR, an intermediate representation, which can be transformed via forward- or reverse-mode AD and compiled by XLA for hybrid architectures. This execution model imposes stronger constraints than NumPy: JAX requires programs to be functional, so mutable state, in-place operations, and unsupported control-flow must be rewritten for tracing compatibility. Thus, the NumPy implementations from previous stages require an additional lowering step to adapt to JAX's model.

FGPT employs a second transformation step that converts NumPy ASTs to JAX. The current implementation produces valid, XLA-compilable \texttt{eqx.Module} subclasses with JAX-traceable methods. However, differentiation interfaces (\texttt{jax.grad}, \texttt{jax.jvp}, \texttt{jax.vjp}) are not yet implemented, as the pipeline does not currently specify which variables are differentiation inputs or outputs. This work therefore focuses on validating the transformation pipeline itself rather than derivative correctness; user-configurable differentiation specifications will be addressed in future work.

JAX transformations (\texttt{jit}, \texttt{grad}, \texttt{jvp}) operate on immutable PyTrees, not Python classes. To preserve object-oriented structure while satisfying JAX's functional model, FGPT uses Equinox \citep{kidger_equinox_2021}, which enables Python classes derived \texttt{eqx.Module} to behave as JAX PyTrees. This allows the NumPy-derived class-based template to be retained without manual restructuring. Equinox is well-suited to our objectives because:

\begin{itemize}
\item Python classes inherit directly from \texttt{eqx.Module}, preserving the object-oriented organization of the translated program;
\item Module instance fields are automatically registered as PyTree leaves, eliminating manual PyTree construction;
\item Static (non-differentiable) fields are identified explicitly using \texttt{eqx.field(static=True)}, allowing metadata such as dimensions and  parameters to remain outside the differentiation graph; 
\item Generated modules remain directly compatible with JAX transformations including \texttt{jit}, \texttt{grad}, \texttt{jvp}, \texttt{vjp}, and \texttt{vmap};
\item Forward and reverse-mode derivatives can be computed directly with respect to differentiable fields of class instances.
\end{itemize}

Compared with higher-level frameworks such as Haiku or Flax, Equinox introduces minimal additional abstraction--a key advantage for automatically generated scientific code that must preserve the structural fidelity of the original Fortran implementation.

\section{JAX pipeline}\label{JAX pipeline}

The JAX lowering step comprises two transformations. First, \texttt{AutoDiff} converts NumPy classes to Equinox modules, establishing the structure required for automatic differentiation. Second, \texttt{JaxConverter} rewrites computational kernels by replacing mutable array updates, adapting control-flow, and substituting NumPy operations with their JAX equivalents. 

\subsection{JaxConverter}\label{jaxconverter}
\subsubsection{Analysis-driven transformation}
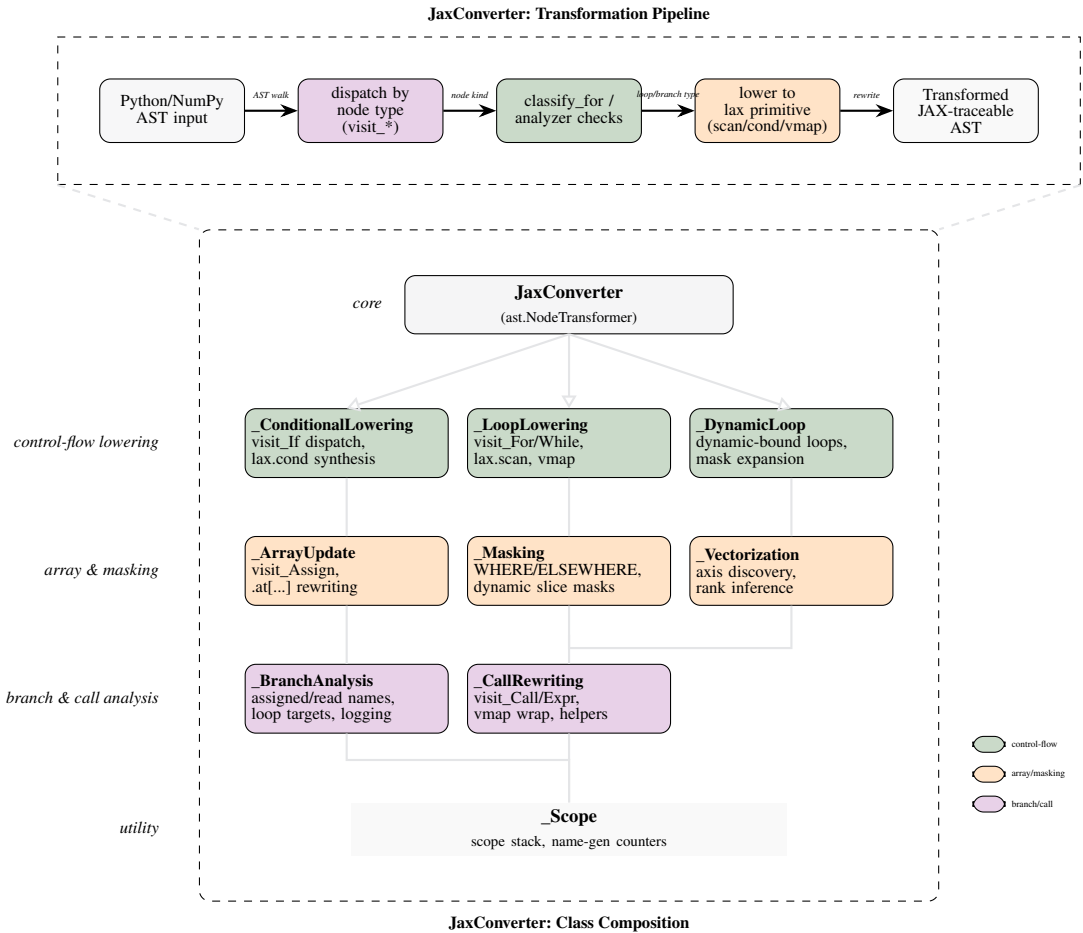
\begin{figure}[t]
\centering
\begin{tikzpicture}[
  scale = 0.7,
  transform shape, 
  node distance=0.9cm and 1.2cm,
  layer/.style={draw, rounded corners, align=center, font=\small},
  core/.style={layer, fill=gray!20, font=\normalsize, minimum width=6.2cm, minimum height=1.1cm},
  pkg/.style={layer, font=\scriptsize, text width=3.6cm, minimum height=1.3cm, align=left},
  ctrlcol/.style={fill=teal!18},
  arrcol/.style={fill=orange!22},
  branchcol/.style={fill=violet!18},
  scopecol/.style={fill=gray!12, text width=8cm, minimum height=0.9cm, align=center},
  layerlabel/.style={font=\scriptsize\itshape, anchor=east},
  inherit/.style={-{Triangle[open]}, thick, gray!55},
  stage/.style={draw, rounded corners, align=center, font=\scriptsize, text width=2.5cm, minimum height=1.2cm},
  flow/.style={-{Stealth}, thick},
  edgelabel/.style={font=\tiny\itshape, above = 0.10cm},
  groupbox1/.style={draw, dashed, inner sep=0pt},
  groupbox/.style={draw, rounded corners, dashed, inner sep=0pt}
]

\node[stage, fill=gray!15] (input) {Python/NumPy\\AST input};
\node[stage, fill=violet!18, right=1.0cm of input] (dispatch) {dispatch by\\node type\\(visit\_*)};
\node[stage, fill=teal!18, right=1.0cm of dispatch] (classify) {classify\_for /\\analyzer checks};
\node[stage, fill=orange!22, right=1.0cm of classify] (lower) {lower to\\lax primitive\\(scan/cond/vmap)};
\node[stage, fill=gray!12, right=1.0cm of lower] (output) {Transformed\\JAX-traceable\\AST};

\draw[flow] (input) -- (dispatch) node[edgelabel, midway] {AST walk};
\draw[flow] (dispatch) -- (classify) node[edgelabel, midway] {node kind};
\draw[flow] (classify) -- (lower) node[edgelabel, midway] {loop/branch type};
\draw[flow] (lower) -- (output) node[edgelabel, midway] {rewrite};

\begin{pgfonlayer}{background}
\node[groupbox1, fit=(input)(dispatch)(classify)(lower)(output),
      inner xsep=0.55cm, inner ysep=0.55cm] (loopgroup) {};
\end{pgfonlayer}

\node[core, below=2.5cm of classify] (core) {\textbf{JaxConverter}\\[1pt] \footnotesize (ast.NodeTransformer)};
\node[layerlabel, left=0.3cm of core] (lbl-core) {core};

\node[pkg, ctrlcol, below=1.4cm of core, xshift=-4.2cm] (cond)
  {\textbf{\_ConditionalLowering}\\visit\_If dispatch,\\lax.cond synthesis};
\node[pkg, ctrlcol, below=1.4cm of core] (loop)
  {\textbf{\_LoopLowering}\\visit\_For/While,\\lax.scan, vmap};
\node[pkg, ctrlcol, below=1.4cm of core, xshift=4.2cm] (dyn)
  {\textbf{\_DynamicLoop}\\dynamic-bound loops,\\mask expansion};
\node[layerlabel, left=1.5cm of cond, xshift=-0cm] (lbl-ctrl) {control-flow lowering};

\node[pkg, arrcol, below=1.1cm of cond] (array)
  {\textbf{\_ArrayUpdate}\\visit\_Assign,\\.at[...] rewriting};
\node[pkg, arrcol, below=1.1cm of loop] (mask)
  {\textbf{\_Masking}\\WHERE/ELSEWHERE,\\dynamic slice masks};
\node[pkg, arrcol, below=1.1cm of dyn] (vec)
  {\textbf{\_Vectorization}\\axis discovery,\\rank inference};
\node[layerlabel, left=1.5cm of array] (lbl-arr) {array \& masking};

\node[pkg, branchcol, below=1.1cm of array] (branch)
  {\textbf{\_BranchAnalysis}\\assigned/read names,\\loop targets, logging};
\node[pkg, branchcol, below=1.1cm of mask] (call)
  {\textbf{\_CallRewriting}\\visit\_Call/Expr,\\vmap wrap, helpers};
\node[layerlabel, left=1.5cm of branch] (lbl-branch) {branch \& call analysis};

\node[scopecol, below=1.3cm of call] (scope)
  {\textbf{\_Scope}\\[1pt]\footnotesize scope stack, name-gen counters};
\node[layerlabel, left=3.5cm of scope] (lbl-scope) {utility};

\foreach \n in {cond, loop, dyn}
  \draw[inherit] (core.south) -- (\n.north);

\draw[thick, gray!55] (cond.south) -- (array.north);
\draw[thick, gray!55] (loop.south) -- (mask.north);
\draw[thick, gray!55] (dyn.south) -- (vec.north);
\draw[thick, gray!55] (array.south) -- (branch.north);
\draw[thick, gray!55] (mask.south) -- (call.north);

\draw[thick, gray!55] (branch.south) -- ++(0,-0.5) -| (scope.north);
\draw[thick, gray!55] (call.south) -- ++(0,-0.5) -| (scope.north);
\coordinate (midMC) at ($(mask.south)!0.5!(call.north)$);

\draw[thick, gray!55]
    (vec.south) -- ++(0,-0.8) -|(midMC);
\node[layer, ctrlcol, right=3.5cm of scope, yshift=1.6cm, font=\tiny, minimum width=0.6cm, minimum height=0.3cm, label={[font=\tiny]right:control-flow}](legend) {};
\node[layer, arrcol, below=0.25cm of legend, font=\tiny, minimum width=0.6cm, minimum height=0.3cm,
label={[font=\tiny]right:array/masking}](arr) {};
\node[layer, branchcol, below=0.25cm of arr, font=\tiny, minimum width=0.6cm, minimum height=0.3cm,
label={[font=\tiny]right:branch/call}](legend3) {}; 

\begin{pgfonlayer}{background}
\node[groupbox,
  fit=(core)(lbl-core)
      (cond)(loop)(dyn)
      (array)(mask)(vec)
      (branch)(call)
      (scope),
  inner xsep=0.6cm, inner ysep=0.6cm] (archgroup) {};
\end{pgfonlayer}

\node[font=\bfseries\small, black, above=0.15cm of loopgroup.north] {JaxConverter: Transformation Pipeline};

\node[font=\bfseries\small, black, below=0.15cm of archgroup.south] {JaxConverter: Class Composition};

\draw[dashed, gray!55, thick] (loopgroup.south west) -- (archgroup.north west);
\draw[dashed, gray!55, thick] (loopgroup.south east) -- (archgroup.north east);

\end{tikzpicture}
\caption{Workflow of the \texttt{JaxConverter} and its helper classes}
\label{fig:jaxconverter_workflow}
\end{figure}

\texttt{JaxConverter} is organized into specialized transformation modules (Figure~\ref{fig:jaxconverter_workflow}) that implement the lowering operations described below. Unlike a direct statement-to-statement translation, this stage is analysis-driven and optimizing: loops and conditionals are first classified by supporting analysis classes--\texttt{VectorizationAnalyser} and \texttt{Control}--which inspect the program structure without modifying the AST and produce metadata describing the appropriate transformation strategy. Specific handling is applied per construct (Table~\ref{tab:control_flow}), as not all are equally traceable.

\begin{table}[t]
    \tabcolsep=8pt%
    \TBL{\caption{Summary of control-flow constructs compatible with \texttt{jit} compilation and automatic differentiation. \texttt{fwd} indicates forward-mode differentiation only; $^{*}$ denotes whether the loop may be unrolled.\label{tab:control_flow}}}
    {\begin{fntable}
    \centering
    \renewcommand{\arraystretch}{1.0}
    \begin{tabular*}{0.8\textwidth}{@{\extracolsep{\fill}} l c c @{}}\toprule
    & \textbf{jit} & \textbf{grad} \\
    \textbf{if}            & \xmark      & \cmark \\
    \textbf{for}           & \cmark$^{*}$ & \cmark \\
    \textbf{while}         & \cmark$^{*}$ & \cmark \\
    \textbf{lax.while\_loop} & \cmark    & \texttt{fwd} \\
    \textbf{lax.fori\_loop}  & \cmark    & \texttt{fwd} \\
    \textbf{lax.scan}       & \cmark      & \cmark \\
    \botrule
    \end{tabular*}%
    \end{fntable}}
\end{table}

\texttt{Control} accumulates the metadata that later stages need to rewrite a given loop or conditional correctly. Conceptually similar to the control stack described in Section~\ref{sec:f2np}, it is designed to carry JAX-specific information: the kind of construct being tracked, its classification (as determined by \texttt{VectorizationAnalyser}), the axis it vectorizes along if applicable, and any additional context needed to preserve nested or masked conditions once the code is rewritten. These decisions--whether to vectorize, apply masked updates, or use indexed loops--guide the \texttt{JaxConverter} in the AST rewrites.

\subsubsection{Vectorization}\label{Vectorization}

Vectorization is applied along a user-specified loop variable provided as a pipeline input rather than inferred automatically. In the \texttt{hydrology} case study, this variable is \texttt{kjpindex}--the dominant and most computationally expensive loop index across most kernels. Instances of this variable are broadcast across dimensions or handled via pairwise indexing to enable parallel execution. Remaining loops are rewritten using \texttt{lax.scan}, which preserves loop structure within the computational graph rather than unrolling it, improving tracing efficiency and reducing memory usage under \texttt{jit}.

Because vectorization introduces additional broadcast dimensions, reduction operations must be handled carefully to recover the correct scalar or lower-dimensional output. \texttt{VectorizationAnalyser} classifies loops and conditionals by vectorization potential, while \texttt{Control} tracks control-flow metadata during transformation. A \texttt{RemoveLogging} pass removes logging and print statements, as these serve no purpose in a differentiable, high-performance context.

\subsubsection{Outer-to-inner strategy}\label{Outer-to-inner strategy}

JAX traces symbolically, propagating abstract tracers (\texttt{shape} and \texttt{dtype}) rather than concrete values, so control-flow and array shapes must be statically known. Data-dependent loops must therefore be restructured into functional primitives such as \texttt{lax.scan} or \texttt{lax.cond} before tracing. This requirement shapes the converter's recursive strategy: rather than a pure depth-first traversal, \texttt{JaxConverter} restructures outer control-flow constructs before recursing into their bodies. For example, converting a \texttt{for} loop into \texttt{lax.scan} first extracts and encapsulates the loop body into a new scan function, and only then applies a depth-first traversal within that encapsulated body via \texttt{self.visit}. This outer-to-inner ordering preserves higher-level semantic context--transforming nested constructs before their enclosing loop or conditional is resolved risks losing vectorization or scan-boundary information.

\subsubsection{Helper-function}\label{Helper-function}

Some JAX control-flow constructs, e.g.,  \texttt{lax.scan} and \texttt{lax.cond}, require their computation bodies to be represented as standalone helper functions. These are constructed incrementally using a worklist during transformation and finalized once the enclosing function is fully processed. Each helper is transformed within its own scoped context, ensuring isolation from sibling and parent scopes.

\subsubsection{Dimensional consistency}

Promoting a variable from a scalar to an array indexed along the vectorization axis may introduce dimensional inconsistencies in expressions that combine it with variables of different shapes. \texttt{MaybeAddIndexTransformer} detects and corrects this situation: it compares the rank each expression is expected to have against the rank it has once vectorization has been applied, and where a lower-rank operand needs to be broadcast against a higher-rank one along the vectorized axis, it inserts the minimal indexing needed to align them. This lets vectorization be applied locally, one variable at a time, without requiring every expression in a procedure to be manually re-derived for shape consistency; only the specific operands that fall out of alignment as a result of vectorization are touched, leaving the rest of the expression unchanged.

\subsubsection{Reduction semantics}

Reduction operations are often written without an explicit axis in the NumPy code, relying on an implicit "reduce over everything" default. This ambiguity is incompatible with JAX tracing and differentiation, where the shape of every intermediate value must be known statically. \texttt{ReductionHandler} resolves this by inferring, for each reduction call, which axis or axes it should act over, based on the structure of the expression being reduced and the known dimensionality of its operands, and then rewriting the call to specify that axis explicitly. Where a reduction's operand carries a vectorized axis, that axis is excluded from the inferred reduction, so that vectorization introduced earlier in the pipeline is not silently collapsed by a later reduction operating over the wrong dimension.

\subsubsection{Corrective passes}

Following the primary transformations, two corrective passes ensure semantic correctness. \texttt{MaybeAddIndexTransformer} resolves shape and broadcasting inconsistencies, while \texttt{ReductionHandler} inserts explicit reduction axes where required to preserve the intended reduction semantics. Like the main transformation modules, these passes operate directly on the AST.

\subsection{AutoDiff}\label{sec:autodiff}

The \texttt{AutoDiff} class orchestrates the transformation of NumPy modules into JAX- and Equinox-compatible programs, structurally preparing them for automatic differentiation in either forward (tangent-linear) or reverse (adjoint) mode. This stage follows the same orchestrated transformation pattern as the Fortran to NumPy pipeline, but must additionally satisfy the functional purity, immutability, and PyTree constraints imposed by JAX and Equinox.

\begin{figure}[h]
    \centering
    \begin{tikzpicture}[
        >=Stealth,
        scale=0.6,
        transform shape,
        process/.style={
            draw,
            rectangle,
            minimum height=1cm,
            minimum width=9cm,
            align=center,
            font=\large,
            line width=0.7pt,
            rounded corners=4pt,
            fill=green!35
        },
        subprocess/.style={
            draw,
            rectangle,
            minimum height=1cm,
            minimum width=8cm,
            align=center,
            font=\normalsize,
            line width=0.7pt,
            rounded corners=4pt,
            fill=red!20
        },
        input/.style={
            draw,
            rectangle,
            minimum height=1cm,
            minimum width=9cm,
            align=center,
            font=\large,
            line width=0.7pt,
            rounded corners=4pt,
            fill=yellow!20
        },
        output/.style={
            draw,
            rectangle,
            minimum height=1cm,
            minimum width=9cm,
            align=center,
            font=\large,
            line width=0.7pt,
            rounded corners=4pt,
            fill=blue!15
        },
        group/.style={
            draw,
            dashed,
            rounded corners,
            inner sep=6pt
        },
        arrow/.style={->, line width=0.9pt},
        node distance=1.2cm
    ]

    \node[input] (input_files)
        {\textbf{Main file} + \textbf{Class file}: parsed and validated as \texttt{ast.Module}};

    \node[process, below=of input_files] (correct_main)
        {\textbf{correct\_main}: patch Main module\\
         (JAX imports, x64, input wrapping)};

    \node[process, fill=orange!30, below=of correct_main] (prepare_class)
        {\textbf{\_prepare\_class}: restructure class module (see right)};

    \node[process, below=of prepare_class] (patch_main)
        {Patch main with \texttt{class\_modif}\\
         (insert timer, fix test calls, rename instances)};

    \node[process, below=of patch_main] (finalize)
        {Fix imports \& write both modules to files};

    \node[output, below=of finalize] (final_output)
        {Final Python files: \texttt{*\_jax.py} / \texttt{*\_d.py} / \texttt{*\_b.py}};

    \draw[arrow] (input_files) -- (correct_main);
    \draw[arrow] (correct_main) -- (prepare_class);
    \draw[arrow] (prepare_class) -- (patch_main);
    \draw[arrow] (patch_main) -- (finalize);
    \draw[arrow] (finalize) -- (final_output);

    \node[subprocess, right=2.8cm of correct_main] (step1)
        {1. Add \texttt{eqx.Module} base, rename class};

    \node[subprocess, below=of step1] (step2)
        {2. Define static / dynamic fields};

    \node[subprocess, below=of step2] (step3)
        {3. Strip numpy scalar casts from \texttt{\_\_init\_\_}};

    \node[subprocess, below=of step3] (step4)
        {4. Rewrite \texttt{declaration\_initialization}\\ for JAX file reading};

    \node[subprocess, below=of step4] (step5)
        {5. Convert method bodies via \texttt{JaxConverter}\\ (topological order, leaves first)};

    \node[subprocess, below=of step5] (step6)
        {6. Decorate outermost method\\ with \texttt{@eqx.filter\_jit}};

    \node[subprocess, below=of step6] (step7)
        {7. Build \texttt{class\_modif} + \texttt{timer\_node}};

    \draw[arrow] (step1) -- (step2);
    \draw[arrow] (step2) -- (step3);
    \draw[arrow] (step3) -- (step4);
    \draw[arrow] (step4) -- (step5);
    \draw[arrow] (step5) -- (step6);
    \draw[arrow] (step6) -- (step7);

    \draw[arrow] (prepare_class.east) -- ++(1cm,0) |- (step1.west);
    \draw[arrow] (step7.west) -| ++(-1.3cm,0) |- (prepare_class.east);
    \node[group,
    fit=(step1)(step2)(step3)(step4)(step5)(step6)(step7)
    ]{};
    \end{tikzpicture}
    \caption{Workflow of \texttt{AutoDiff}}
    \label{fig:autodiff_workflow}
\end{figure}

\subsubsection{Ordering }\label{Ordering }

Unlike the Fortran to NumPy transpilation, which proceeds sequentially from class definitions to calling scripts, the Equinox transformation begins with the \texttt{main} script. This ordering is necessary because \texttt{main} defines the concrete input arguments (e.g., dimensions and types) that seed shape propagation and guide subsequent transformations--information required both for the NumPy to JAX  conversion and for correct tracing downstream.

The \texttt{main} function is rewritten to enable JAX double precision, replace NumPy arrays with \texttt{jnp.asarray}, explicitly assign JAX dtypes (\texttt{jnp.int32}, \texttt{jnp.float64}) to scalar inputs, and insert two calls to the transformed function: a warm-up call that triggers JAX tracing and compilation, and a second call that measures execution performance with compilation overhead excluded. Additional script-level changes insert the required JAX and Equinox imports, convert \texttt{np} operations to \texttt{jnp}, and integrate a timing utility. As with earlier stages, all modifications are performed on the AST and the scripts are regenerated via \texttt{ast.unparse}.

Class definitions are then rewritten to inherit from \texttt{eqx.Module}, converting them into Equinox PyTree modules. Fields are partitioned into {dynamic fields} (differentiable arrays and runtime values) and {static fields}, marked via \texttt{eqx.field(static=True)} to exclude structural parameters from differentiation. This separation is required because Equinox needs explicit control over which attributes participate in PyTree traversal, and by extension in any future gradient computation.

\subsubsection{Functionalizing mutations}\label{Functionalizing mutations}

Beyond syntactic \texttt{np}-to-\texttt{jnp} substitution, the class-level \texttt{read} method requires further restructuring because JAX arrays are immutable and Equinox modules behave as PyTrees--in-place attribute updates of the form \texttt{self.attribute = value} are not permitted. Assignments are therefore rewritten into a functional update scheme: scalar values are wrapped with explicit JAX dtypes, array initializations (e.g., \texttt{np.zeros}) are adjusted to JAX constructors, and direct assignments or \texttt{setattr} calls are replaced with updates to an intermediate dictionary that accumulates modified attributes. Once all assignments are processed, the model instance is functionally reconstructed via \texttt{eqx.tree\_at}, producing a new PyTree with the updated fields rather than mutating the original object. This makes the resulting module structurally compatible with JAX transformations such as \texttt{jit}, which require side-effect-free execution.

Trivial control-flow branches that could introduce data-dependent execution paths, e.g., conditionals containing only a \texttt{continue} statement, are removed, as they could interfere with symbolic tracing.

\subsubsection{Assembly}\label{Assembly}

Once structural and functional transformations are complete, the class's computational methods are processed by \texttt{JaxConverter} (Section~\ref{jaxconverter}), transformed in dependency order so that callees are rewritten before their callers. The parent function is decorated with \texttt{eqx.filter\_jit} and inserted back into the class definition.

The output of this stage is a fully functional Equinox module that preserves the semantics of the original NumPy implementation while satisfying JAX's tracing requirements. Figure~\ref{fig:autodiff_workflow} illustrates this workflow, showing the transformation of the \texttt{main} module and class module, including functional updates, JAX conversion, and final code generation via \texttt{ast.unparse}.

\begin{figure}[t]
    \centering
    \begin{subfigure}[b]{0.45\textwidth}
        \centering
        \includegraphics[width=\textwidth]{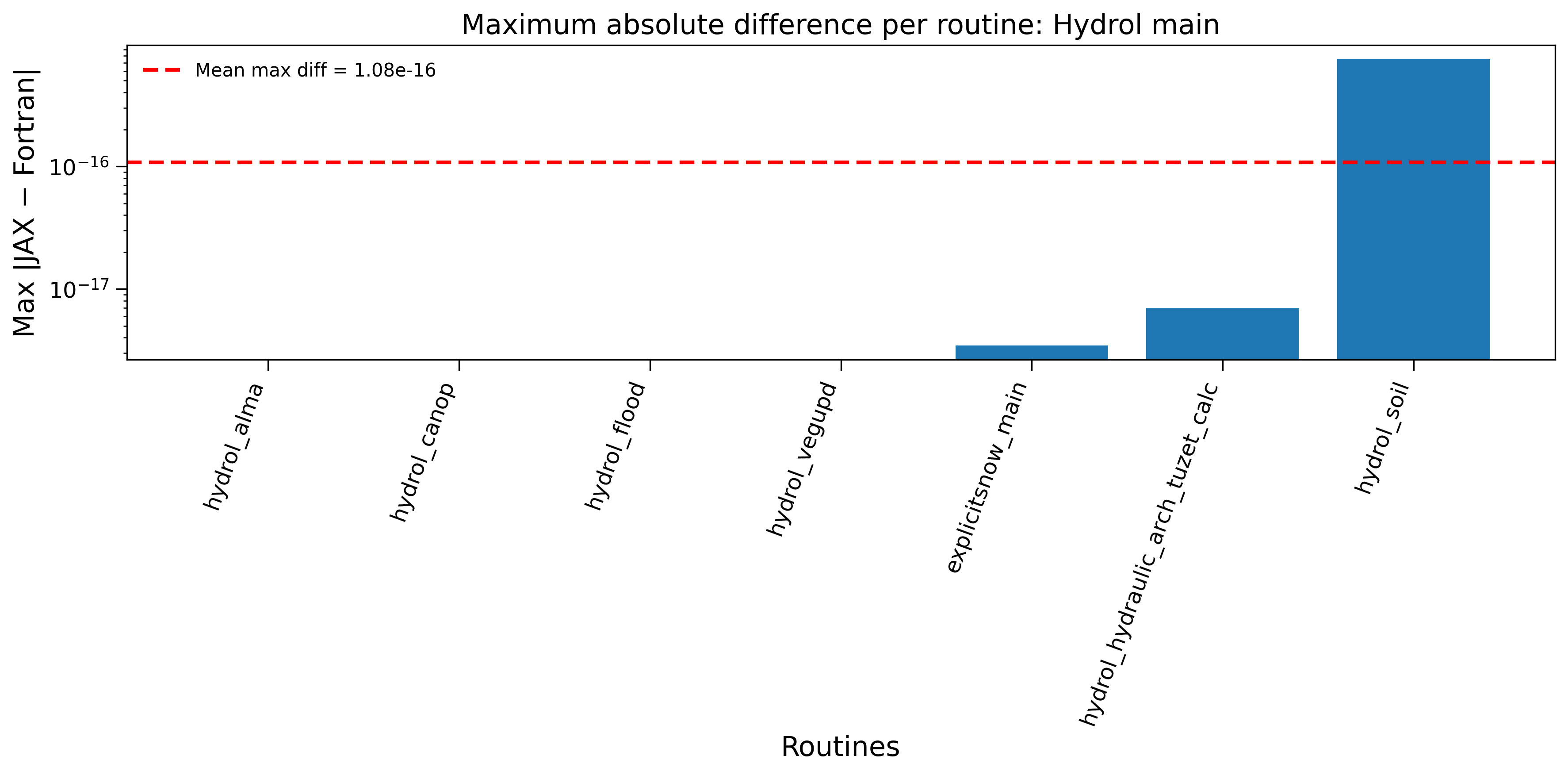}
        \caption{\texttt{hydrol\_main}.}
        \label{fig:absdiff_jax_hydrol_main}
    \end{subfigure}
    \hfill
    \begin{subfigure}[b]{0.45\textwidth}
        \centering
        \includegraphics[width=\textwidth]{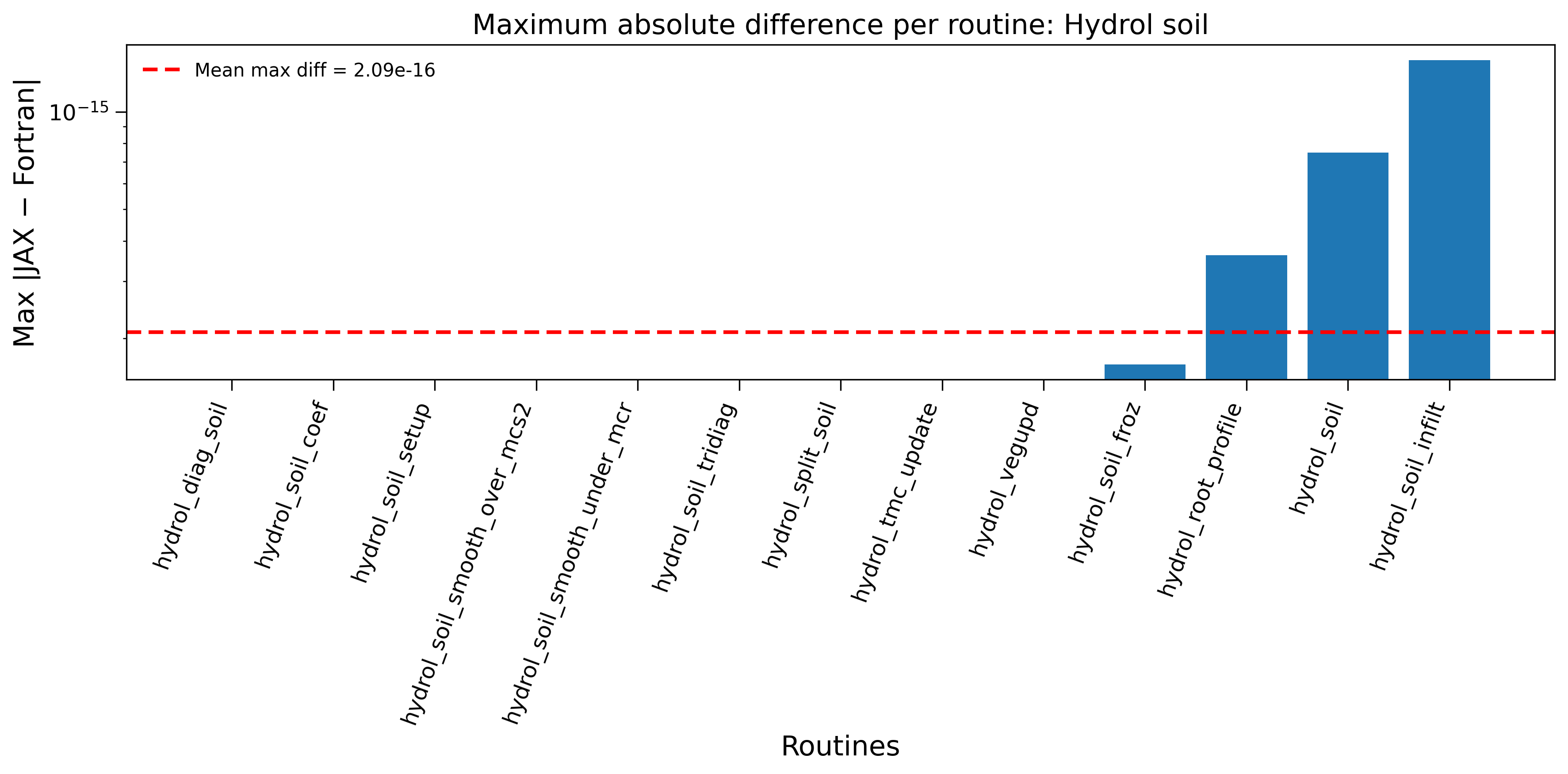}
        \caption{\texttt{hydrol\_soil}.}
        \label{fig:absdiff_jax_hydrol_soil}
    \end{subfigure}
    \vskip\baselineskip
    \begin{subfigure}[b]{0.45\textwidth}
        \centering
        \includegraphics[width=\textwidth]{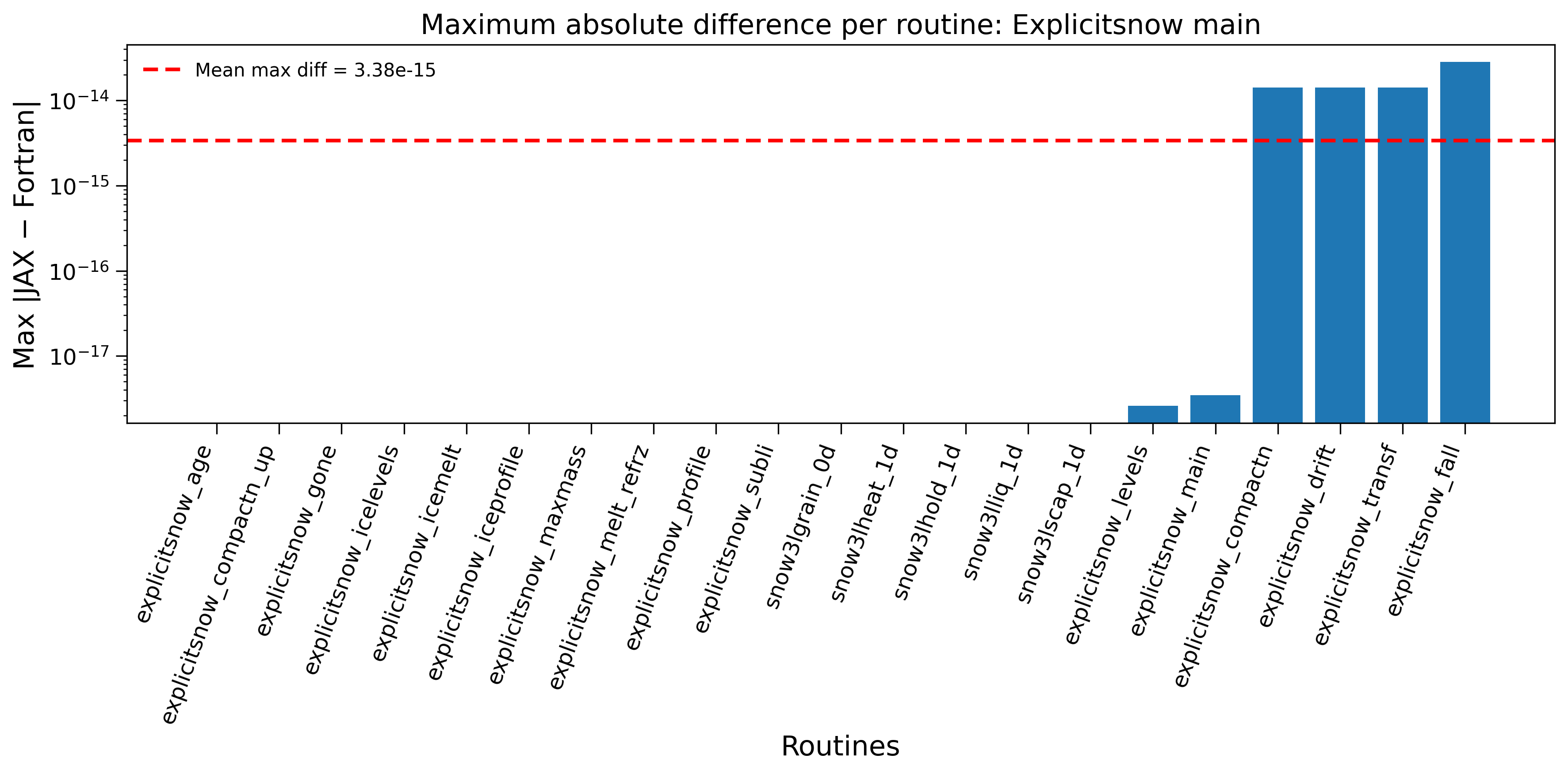}
        \caption{\texttt{explicitsnow\_main}.}
        \label{fig:absdiff_jax_explicitsnow_main}
    \end{subfigure}
    \hfill
    \begin{subfigure}[b]{0.45\textwidth}
        \centering
        \includegraphics[width=\textwidth]{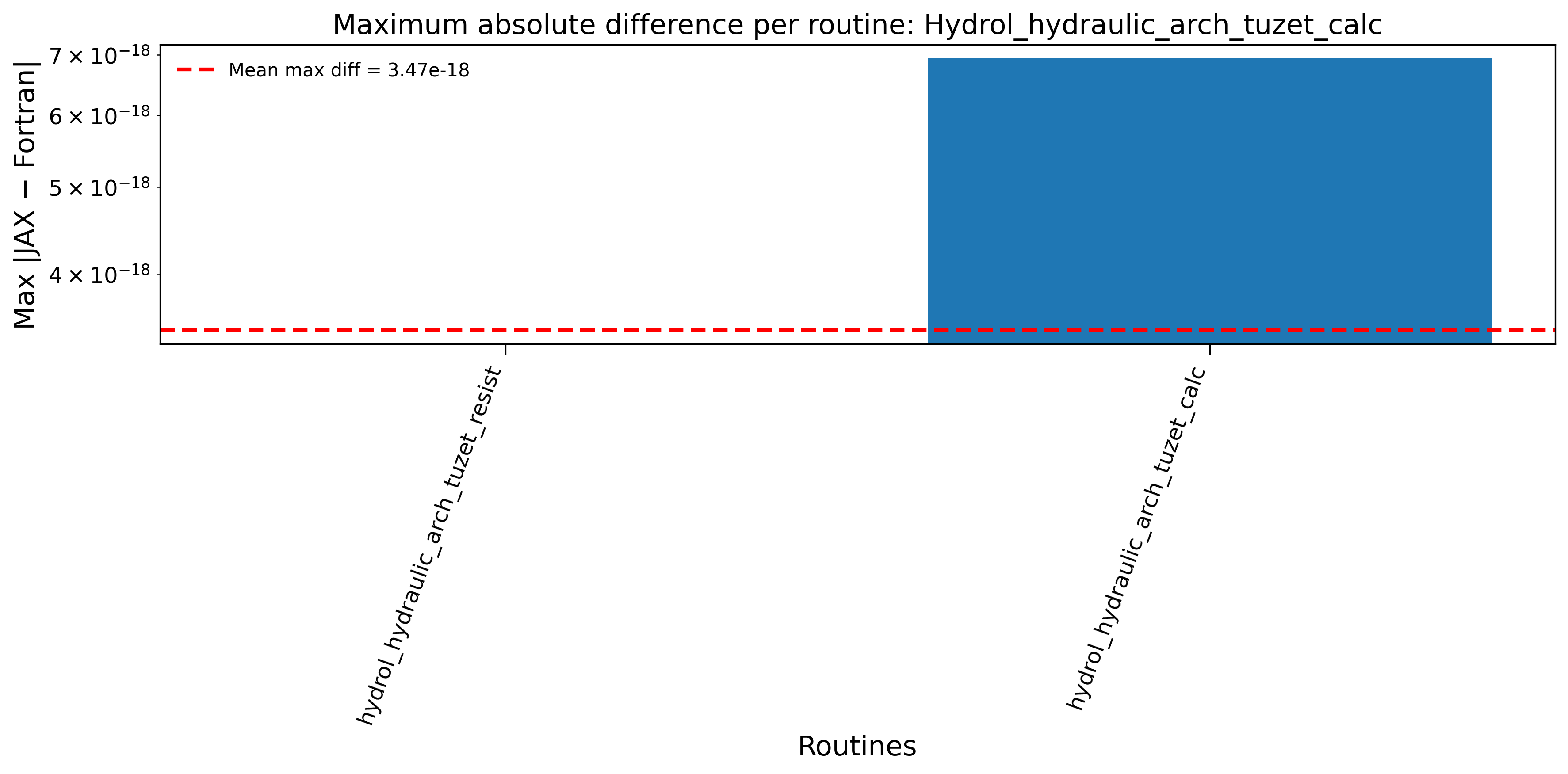}
        \caption{\texttt{hydrol\_hydraulic\_arch\_tuzet\_calc}.}
        \label{fig:absdiff_jax_hydrol_hydraulic_calc}
    \end{subfigure}

    \caption{Maximum absolute deviation between JAX and Fortran outputs}
    \label{fig:test_comp_jax}
\end{figure}

\subsection{Numerical verification}\label{sec:jax_validation}

This section extends the verification framework from Section~\ref{sec:validation} to the JAX implementations, assessing both numerical consistency and performance. The same validation protocol--maximum absolute difference and wall-clock timing--is applied throughout. 

As shown in Figure~\ref{fig:test_comp_jax}, numerical consistency is preserved throughout the complete translation pipeline, from Fortran to Python and subsequently to JAX. Most routines fall into one of two groups: exact matches, or exhibit only floating-point differences on the order of \(10^{-14}\) to \(10^{-18}\) for the highest-level parent routines (e.g., \texttt{explicitsnow\_main}, and \texttt{hydrol\_soil}, where small floating-point rounding differences accumulate along the call chain. All observed differences remain several orders of magnitude below the validation tolerances (\texttt{rtol=1e-5}, \texttt{atol=1e-8}).

\begin{figure}[htbp]
    \centering
    \begin{subfigure}[b]{0.45\textwidth}
        \centering
        \includegraphics[width=\textwidth]{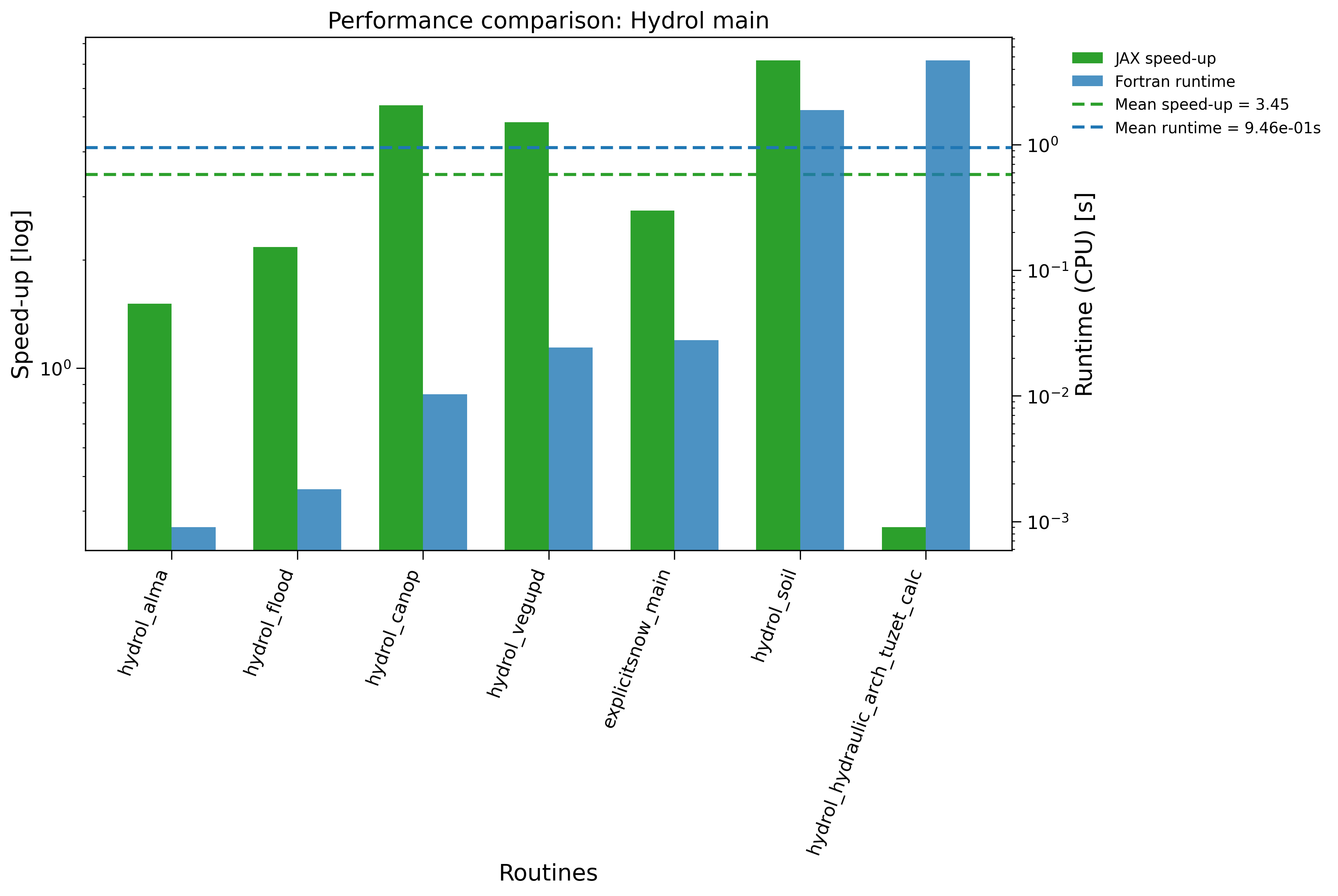}
        \caption{\texttt{hydrol\_main}.}
        \label{fig:perf_jax_hydrol_main}
    \end{subfigure}
    \hfill
    \begin{subfigure}[b]{0.45\textwidth}
        \centering
        \includegraphics[width=\textwidth]{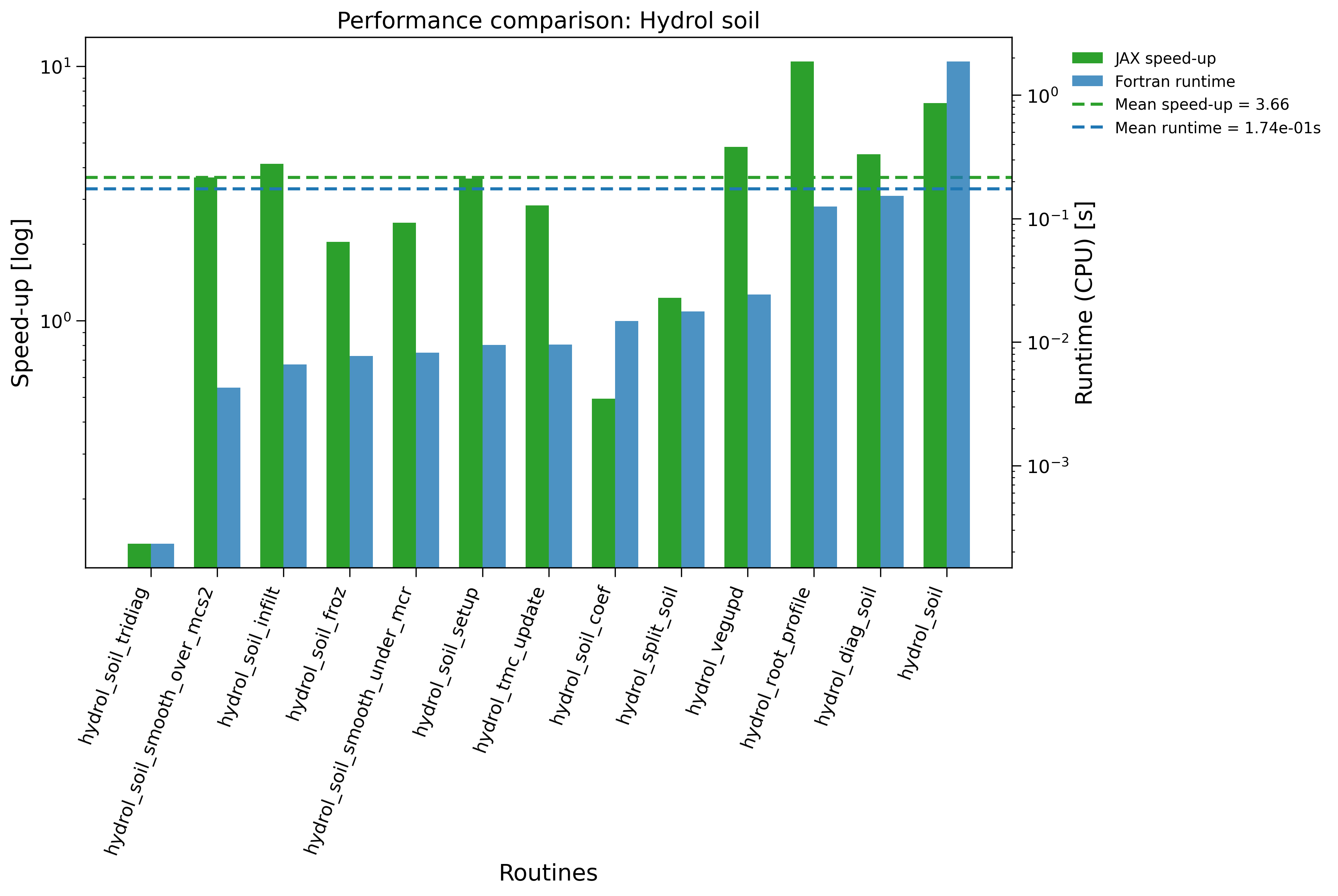}
        \caption{\texttt{hydrol\_soil}.}
        \label{fig:perf_jax_hydrol_soil}
    \end{subfigure}
    \vskip\baselineskip
    \begin{subfigure}[b]{0.45\textwidth}
        \centering
        \includegraphics[width=\textwidth]{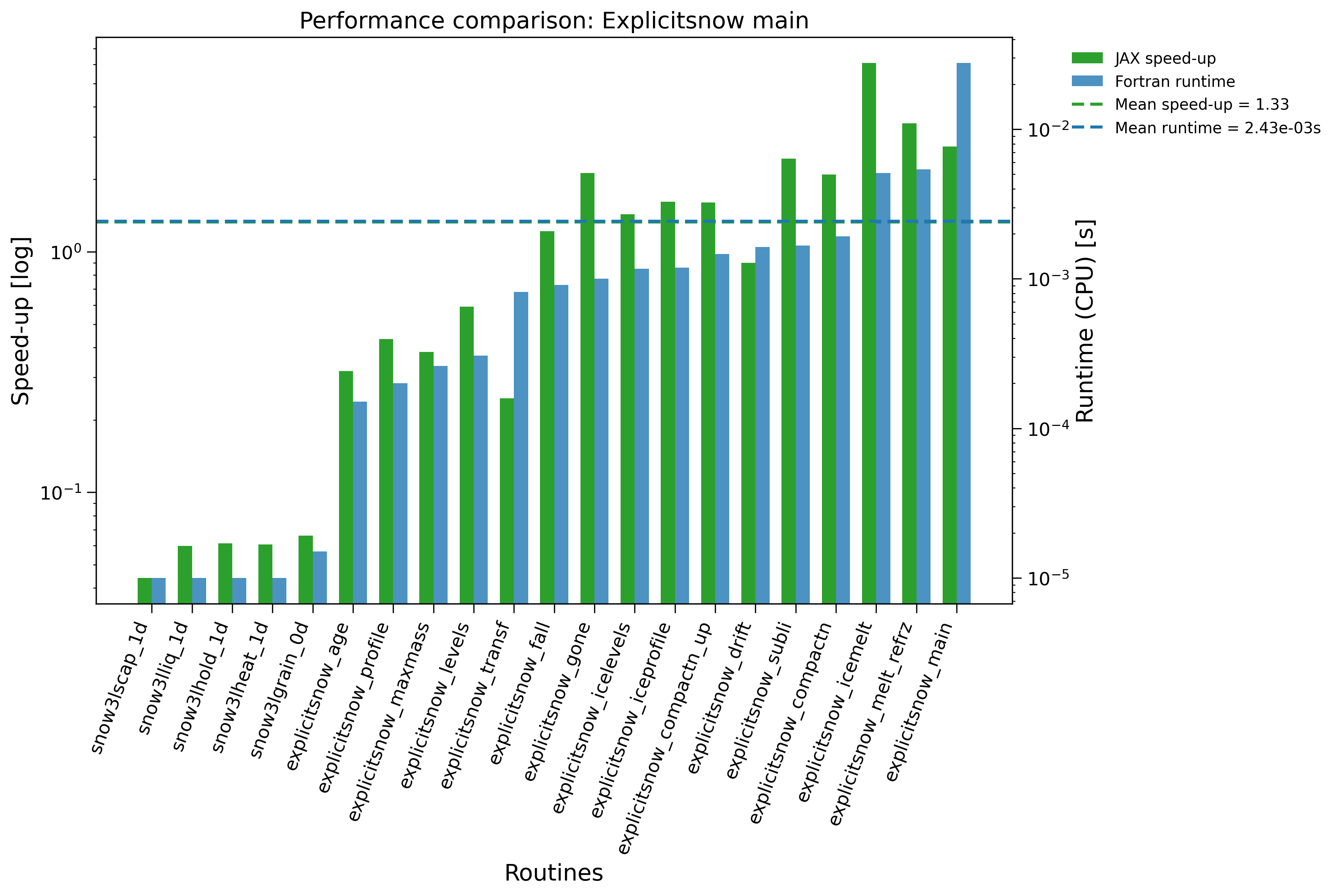}
        \caption{\texttt{explicitsnow\_main}.}
        \label{fig:perf_jax_explicitsnow_main}
    \end{subfigure}
    \hfill
    \begin{subfigure}[b]{0.45\textwidth}
        \centering
        \includegraphics[width=\textwidth]{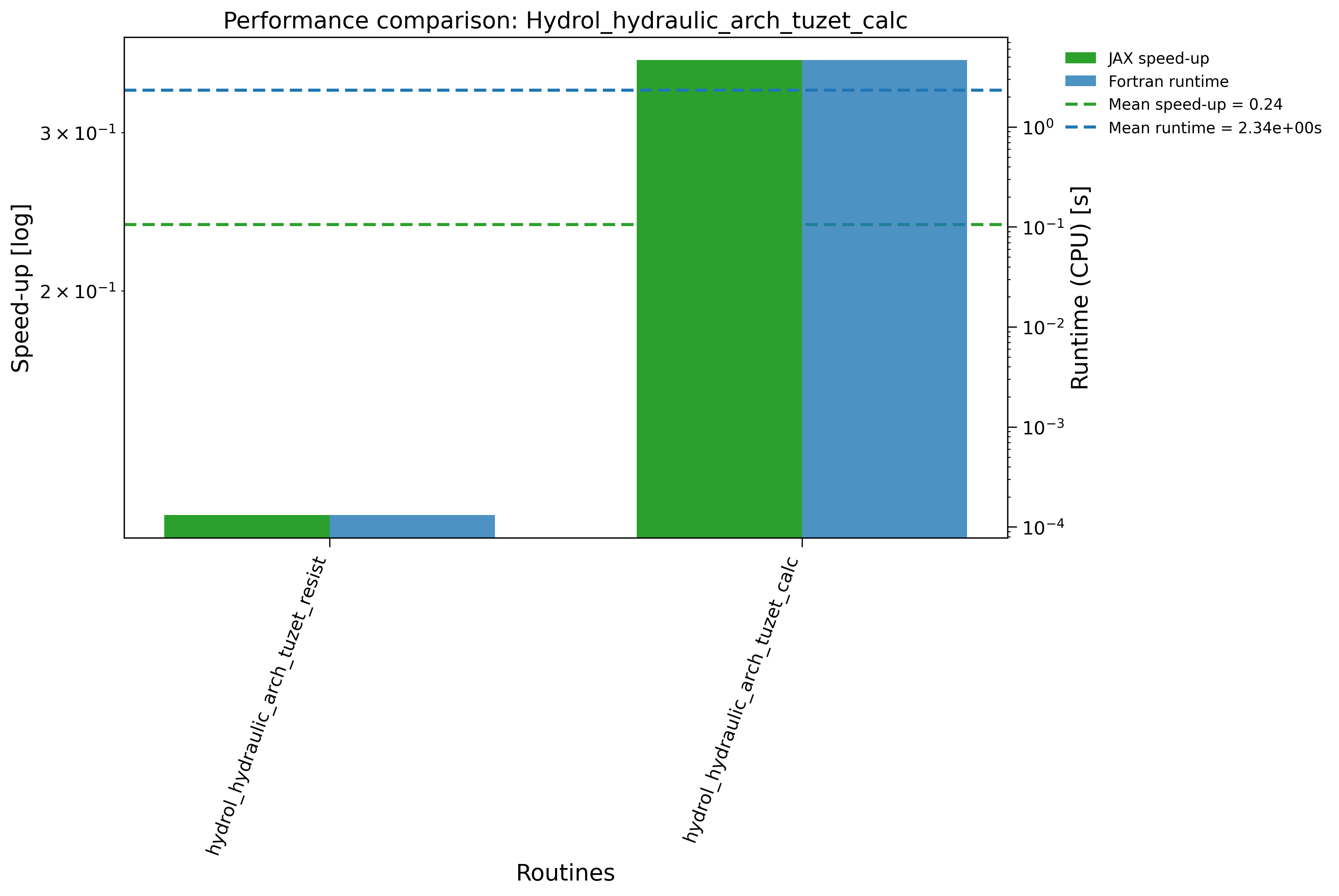}
        \caption{\texttt{hydrol\_hydraulic\_arch\_tuzet\_calc}.}
        \label{fig:perf_jax_hydrol_hydraulic_calc}
    \end{subfigure}

    \caption{Runtime comparison and speedup between Jax and Fortran}
    \label{fig:speed_up_jax}
\end{figure}

\begin{figure}[b]
    \centering
    \includegraphics[width= 0.6\textwidth]{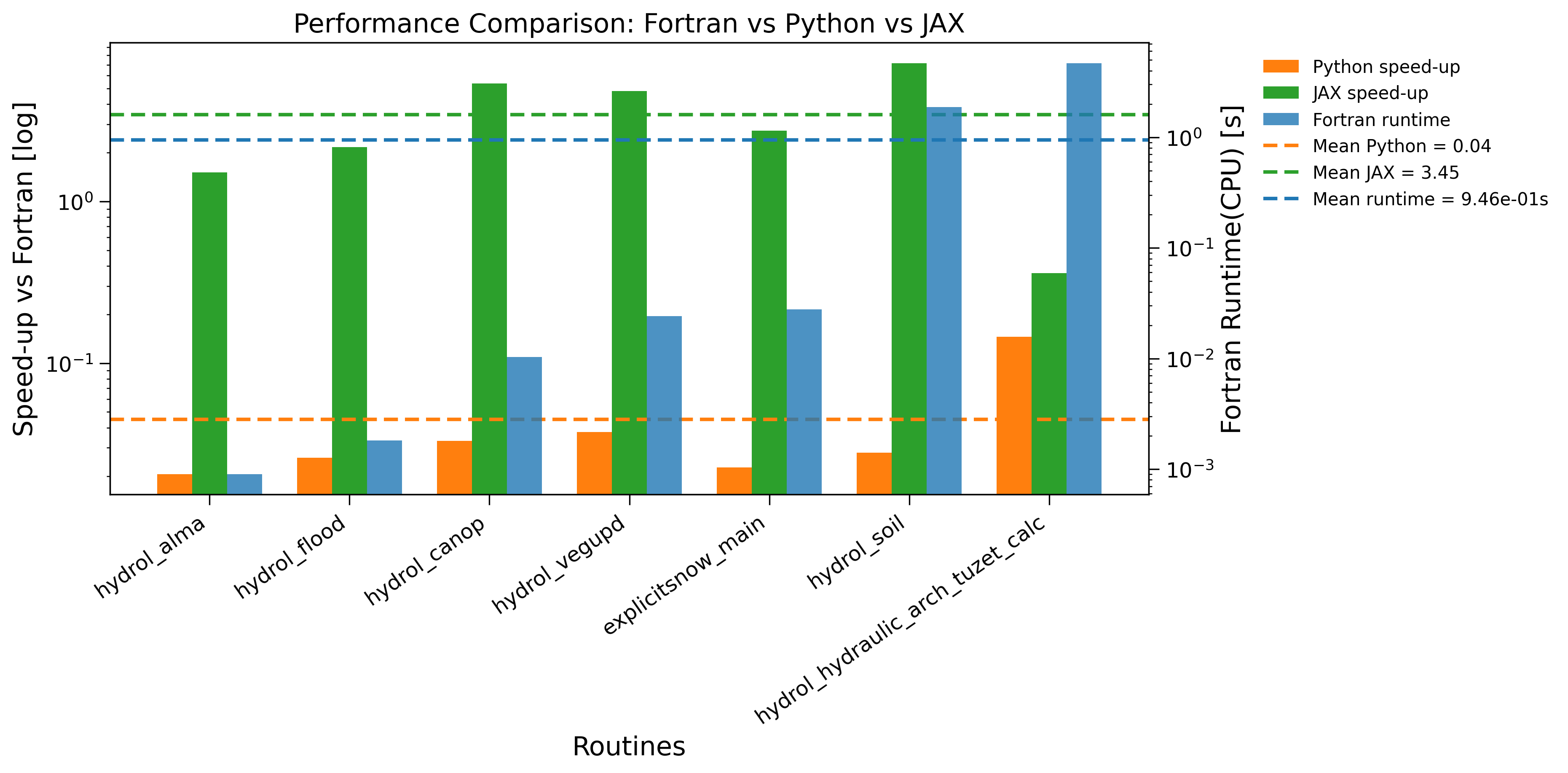}
    \caption{Performance evaluation between Fortran, NumPy and JAX}
    \label{fig:f_py_jax_comparaison}
\end{figure}

Performance is evaluated for all three implementations: original compiled Fortran, NumPy, and JIT-compiled JAX. Figure~\ref{fig:perf_jax_hydrol_main} reports execution time and speedup for \texttt{hydrology}, while Figures~\ref{fig:perf_jax_hydrol_soil}--\ref{fig:perf_jax_hydrol_hydraulic_calc} present the corresponding results for its child procedures. Figure~\ref{fig:f_py_jax_comparaison} then summarizes all three configurations together, reporting both NumPy and JAX speedups relative to Fortran alongside the Fortran runtime for each high-level routine.

Across nearly all procedures, the JAX implementation is faster than both the Fortran reference and the NumPy implementation. This speedup is consistent throughout the procedure hierarchy--it holds for both parent routines and their lower-level children, indicating that the benefits of vectorization and JIT compilation propagate across the call chain. As shown in Figure~\ref{fig:f_py_jax_comparaison}, the JIT-compiled JAX implementation achieves a 3.66$\times$ speedup over the Fortran implementation, owing to vectorization and JIT compilation. This improvement is particularly evident for routines such as \texttt{hydrol\_soil}, which exhibits significantly faster execution (Table~\ref{tab:runtime_values}).

\begin{table*}[t]
\tabcolsep=3pt%
\TBL{\caption{JAX, NumPy, and Fortran runtimes in seconds.\label{tab:runtime_values}}}
{\begin{fntable}
\begin{tabular*}{\textwidth}{@{\extracolsep{\fill}} l *{7}{c} @{}}\toprule
 & \rotatebox{45}{\textbf{hydrol\_alma}}
 & \rotatebox{45}{\textbf{hydrol\_canop}}
 & \rotatebox{45}{\textbf{hydrol\_flood}}
 & \rotatebox{45}{\textbf{hydrol\_vegupd}}
 & \rotatebox{45}{\textbf{hydrol\_soil}}
 & \rotatebox{45}{\textbf{explicitsnow\_main}}
 & \rotatebox{45}{\textbf{hydrol\_hydraulic\_arch\_tuzet\_calc}} \\
\textbf{JAX(CPU)}
      & \textbf{0.000599}
      & \textbf{0.001840}
      & \textbf{0.000796}
      & \textbf{0.004448}
      & \textbf{0.257696}
      & \textbf{0.009416}
      & 13.007017 \\
      
\textbf{NumPy(CPU)}
      & 0.043769
      & 0.308702
      & 0.068819
      & 0.649055
      & 66.780735
      & 1.222364
      & 32.056070 \\
      
\textbf{Fortran(CPU)}
      & 0.000904
      & 0.010225
      & 0.001814
      & 0.023951
      & 1.960143
      & 0.027354
      & \textbf{4.606041} \\
\botrule
\end{tabular*}%
\end{fntable}}
\end{table*}

The only exception is \texttt{hydrol\_hydraulic\_arch\_tuzet\_calc}, whose Fortran implementation consists of an outer vectorization loop that repeatedly calls child routines--a structure that cannot be vectorized by JAX, resulting in no performance gain over the Fortran version.

\subsubsection{NumPy and JAX numerical consistency}
Although the Python and JAX implementations are generated from the same intermediate representation, they do not produce identical numerical differences with respect to the Fortran. This is expected: the two backends execute the same mathematical expressions through different execution stacks, causing floating-point rounding errors to accumulate independently. The JAX implementation is compiled through XLA, which may fuse, reorder, or vectorize arithmetic operations, producing a different sequence of floating-point operations and, consequently, slightly different rounding behavior.

For example, in the routine \texttt{explicitsnow\_main}, the mean maximum absolute difference is \(\sim 10^{-12}\) for the NumPy implementation and \(\sim 10^{-16}\) for the JAX implementation. For both backends, the largest numerical deviations occur in the same parent routines (e.g., \texttt{explicitsnow\_main, hydrol\_soil}), indicating that the dominant source of numerical discrepancy is the accumulation of rounding error along the call hierarchy, rather than any backend-specific implementation difference.

\subsection{AD verification}
The current pipeline produces JAX and Equinox modules that are structurally ready for differentiation but does not yet produce \texttt{jax.grad}, \texttt{jax.jvp}, or \texttt{jax.vjp} call sites, since the pipeline does not yet specify which variables gradients should be taken with respect to. Therefore, gradient evaluation is outside the scope of the present work. Nevertheless, the generated modules are fully compatible with JAX's transformation and are structurally prepared for AD. Future AD extensions will also include jnp.where-based safeguards for numerically sensitive operations. Once differentiation targets are introduced, gradient correctness will be validated by comparing JAX-derived gradients against those produced by Tapenade from the original Fortran implementation.

\section[Conclusions]{Conclusions}
This work presented FGPT, a general-purpose source-to-source transpilation framework for modernizing legacy Fortran scientific code through automatic translation into NumPy and JAX implementations. FGPT combines dependency-aware procedure isolation with an AST-based transformation pipeline that preserves the semantic structure of the original program while producing readable, maintainable Python code suitable for scientific computing.

The framework is organized around two transformation stages. The \texttt{F2NP} component performs the semantic lowering of Fortran codes into equivalent NumPy implementations, resolving language-specific differences such as array indexing, intrinsic procedures, control-flow reconstruction, and procedure interfaces. The \texttt{Transformer} component assembles these translated procedures into complete executable Python modules by resolving dependencies, constructing the global program structure, generating initialization logic, and producing standalone programs for execution and verification. A subsequent  stage corrects NumPy indexing semantics and resolves object-oriented attribute mappings.

Building upon the NumPy representation, FGPT further generates JAX implementations by transforming the intermediate Python AST into executable \texttt{eqx.Module} classes and adapting language constructs to satisfy JAX's tracing and functional programming model. Mutable array operations and incompatible control-flow constructs are rewritten using JAX primitives such as \texttt{lax.scan}, \texttt{jax.less\_than}, and \texttt{lax.cond}, producing code that is compatible with XLA compilation and accelerators.

Verification on representative modules from the IPSL land-surface model demonstrates that the framework preserves numerical consistency to within floating-point precision while successfully translating complex scientific code containing deeply nested procedure hierarchies, multidimensional arrays, and extensive module dependencies. These results demonstrate that a compiler-inspired, dependency-aware transpilation pipeline provides a practical and scalable alternative to manual rewriting and emerging LLM-assisted translation approaches by offering deterministic, reproducible transformations.

The current implementation nevertheless has several limitations. Support for all modern Fortran language features has not yet been fully implemented; Table~\ref{tab:coverage} summarizes the constructs currently handled by FGPT alongside those identified for future extension. Additional work is required to extend the translation rules to these more advanced language constructs and object-oriented features. Future releases will therefore expand the coverage of the transpiler while also introducing configurable AD interfaces capable of generating \texttt{jax.grad}, \texttt{jax.jvp}, and \texttt{jax.vjp} wrappers directly from user-specified differentiation targets. Together, these extensions will broaden the applicability of FGPT to increasingly complex scientific codes and further facilitate the modernization of legacy Earth system and high-performance computing applications.

\begin{table}[htbp]
    \tabcolsep=8pt%
    \TBL{\caption{Fortran language constructs in FGPT.\label{tab:coverage}}}
    {\begin{fntable}
    \centering
    \renewcommand{\arraystretch}{1.0}
    \begin{tabular*}{0.8\textwidth}{@{\extracolsep{\fill}} l c @{}}\toprule
    \textbf{Fortran Construct} & \textbf{Status} \\
    DO loops & Supported \\
    IF / ELSE IF / ELSE & Supported \\
    WHERE / ELSEWHERE & Supported \\
    CASE / SELECT CASE & Supported \\
    Array declarations and slicing & Supported \\
    Intrinsic procedures (SUM, MIN, MAX, etc.) & Supported \\
    Module variables and shared state & Supported \\
    External subroutine/function calls & Supported \\
    Multidimensional array operations & Supported \\
    Pointers & Not yet supported \\
    Derived Types & Not yet supported \\
    Interface Blocks & Not yet supported \\
    \botrule
    \end{tabular*}%
    \end{fntable}}
\end{table}

\begin{Backmatter}

\paragraph{Acknowledgments}
We acknowledge the EuroHPC Joint Undertaking for awarding access to the EuroHPC supercomputer LEONARDO, hosted by CINECA (Italy) and the LEONARDO consortium, through the EuroHPC Development Access calls (EHPC-DEV-2024D10-070 and EHPC-DEV-2025D08-095). We also express our sincere gratitude to Sven Rodriguez, Redouane Lguensat, Pascal Maugis, Frederic Chevallier, Fabienne Maignan, and Philippe Peylin for their valuable discussions and insightful suggestions.

\paragraph{Funding Statement}
This work was supported by the Horizon Europe project AI4PEX (Grant No. 101137682).

\paragraph{Competing Interests}
None

\paragraph{Data Availability Statement}
The FGPT source code is publicly available at \href{https://github.com/kardaneh/Fgpt}{https://github.com/kardaneh/Fgpt} under the Creative Commons Attribution-NonCommercial-ShareAlike 4.0 International License. The benchmark data, including the isolated Fortran procedures, transpiled NumPy and JAX implementations, and reference inputs and outputs used for validation, are available on Zenodo at \href{https://zenodo.org/records/21640037}{https://zenodo.org/records/21640037}. The IPSL land-surface model used for evaluation is available to the research community under the terms of its own license.

\paragraph{Ethical Standards}
The research meets all ethical guidelines, including adherence to the legal requirements of the study country.

\paragraph{Author Contributions}
All authors contributed equally.

\paragraph{Supplementary Material}
None

\end{Backmatter}

\newpage
\begin{appendix}\appheader
\section{Appendix}\label{appendixA}
\renewcommand{\thefigure}{A\arabic{figure}}  
\setcounter{figure}{0}  
\renewcommand{\thetable}{A\arabic{table}}
\setcounter{table}{0}

The Fortran codes were compiled using \texttt{mpif90} with the NVIDIA HPC SDK \texttt{nvfortran} backend. CPU builds used the following flags:
\begin{verbatim}
-Wall -g -O0 -Kieee -Ktrap=fp -Mbounds -traceback -r8 -i4
\end{verbatim}
The \texttt{-r8} flag promotes all default \texttt{REAL} variables to 8-byte precision, while \texttt{-i4} sets default \texttt{INTEGER} to 4 bytes. GPU builds additionally used \texttt{-acc -gpu=cc80}, though GPU execution was not used for the benchmarks reported here. The Fortran environment linked against NetCDF (C and Fortran), IOIPSL, XIOS, and Tapenade.

All benchmarks were executed on a single core of an AMD EPYC 7302 16-Core Processor with 6 GiB of RAM, running Ubuntu 20.04.6 LTS. The Python environment consisted of Python 3.10.20, NumPy 2.2.6, and JAX 0.6.2.

\begin{figure}[t]
\centering
\begin{tikzpicture}[
    >=Stealth,
    scale = 0.65,
    transform shape, 
    node distance=1.1cm and 1.6cm,
    every node/.style={font=\small},
    stage/.style={
        draw=black!70,
        rectangle,
        rounded corners=3pt,
        minimum height=1cm,
        minimum width=5cm,
        align=center,
        line width=0.8pt,
        fill=blue!12
    },
    loopbox/.style={
        draw=black!70,
        rectangle,
        rounded corners=3pt,
        minimum height=0.9cm,
        minimum width=5.6cm,
        align=center,
        line width=0.7pt,
        fill=red!10
    },
    groupbox/.style={draw=black!50, rectangle, rounded corners=6pt, line width=0.7pt, dashed},
    source/.style={stage, fill=yellow!25, line width=1pt},
    final/.style={stage, fill=orange!30, line width=1pt},
    pass/.style={stage, fill=green!25, minimum height=0.9cm},
    branch/.style={stage, minimum width=4.2cm, fill=blue!12},
    arrow/.style={->, line width=0.9pt, black!70},
    lbl/.style={font=\scriptsize\itshape, black!60},
    dashed arr/.style={-{Stealth[length=5pt]}, dashed, gray!60},
]

\node[source] (fortran) {Fortran Source Code};
\node[loopbox, below=1.6cm of fortran] (loop)
{\textbf{Isolation and Extraction}\\[1pt]
{\scriptsize Parsing, normalization, and procedure-unit extraction.}};

\node[groupbox, fit=(loop), inner xsep=0.55cm, inner ysep=0.55cm] (loopgroup) {};
\node[pass, below=1.3cm of loopgroup] (f2np)
    {F2NP Transformation\\[1pt]
     {\scriptsize Fortran Procedure $\rightarrow$ Python Function}};

\node[branch, below left=1.5cm and -0.6cm of f2np] (global)
    {Global Procedure\\Extraction};
\node[branch, below right=1.5cm and -0.6cm of f2np] (main)
    {Main Procedure\\Extraction};

\node[stage, below=of global] (class_emit) {Python Class Emission};
\node[stage, below=of main] (script_emit) {Python Script Emission};

\coordinate (mid) at ($(class_emit.south)!0.5!(script_emit.south)$);
\node[source, below=1.4cm of mid] (py_src) {Generated Python Source (AST)};

\node[pass, below=of py_src] (rg)
    {ReplaceGlobals\\{\scriptsize Name and Attribute resolution}};
\node[pass, below=of rg] (ai)
    {AdjustIndices\\{\scriptsize Index Normalization}};
\node[pass, below=of ai] (dep) {Dependency Ordering};

\node[final, below=of dep] (finalnode) {Final Normalized Python Code};

\draw[arrow] (fortran) -- (loopgroup);
\draw[arrow] (loopgroup) -- node[lbl, pos=0.5, left] {Fortran AST} (f2np);

\coordinate (split) at ($(f2np.south)+(0,-0.8)$);
\draw[-] (f2np.south) -- node[lbl, pos=0.5, left] {Python AST} (split);
\draw[arrow] (split) -| node[lbl, pos=0.5, above] {Global scope} (global.north);
\draw[arrow] (split) -| node[lbl, pos=0.5, above] {Main routine} (main.north);

\draw[arrow] (global) -- (class_emit);
\draw[arrow] (main) -- (script_emit);

\draw[arrow] (class_emit.south) |- (py_src.west);
\draw[arrow] (script_emit.south) |- (py_src.east);

\draw[arrow] (py_src) -- (rg);
\draw[arrow] (rg) -- (ai);
\draw[arrow] (ai) -- (dep);
\draw[arrow] (dep) -- (finalnode);

\end{tikzpicture}

\caption{End-to-end transpilation pipeline}
\label{fig:full_pipeline}
\end{figure}
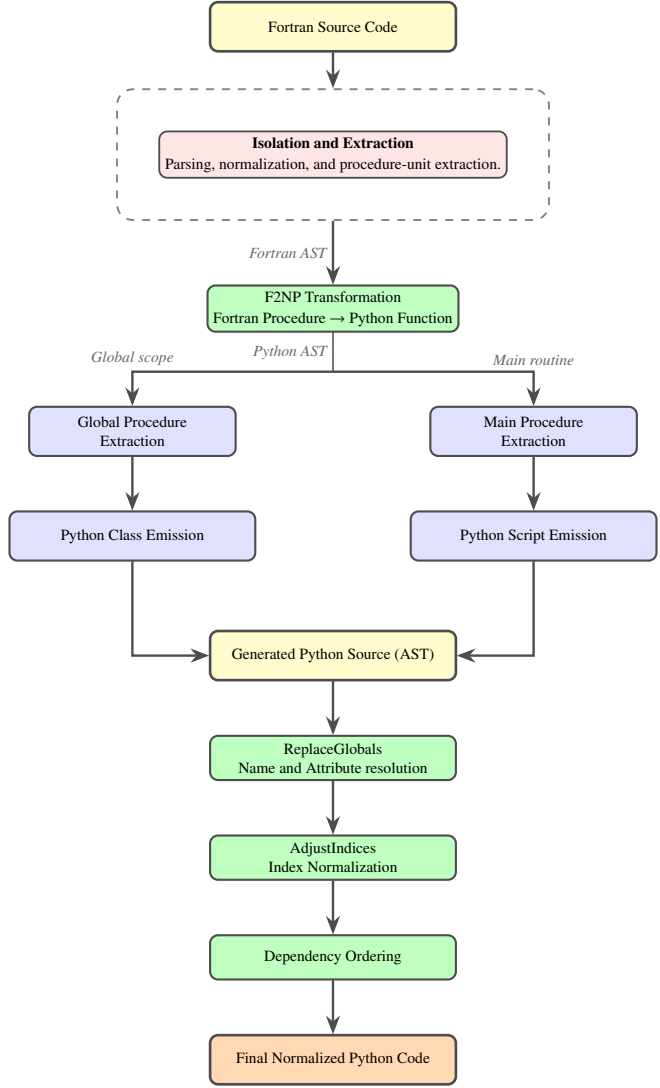

\begin{figure}[t]
    \centering
    \begin{tikzpicture}[
        node distance=1.2cm and 1.2cm,
        scale = 0.5,
        transform shape, 
        >=Stealth,
        terminal/.style={
            draw,
            rounded corners=8pt,
            minimum height=0.9cm,
            minimum width=3cm,
            align=center,
            font=\large\bfseries,
            line width=0.8pt,
            fill=blue!10
        },
        state/.style={
            draw,
            diamond,
            minimum width=2cm,
            minimum height=2cm,
            align=center,
            font=\large\bfseries,
            line width=0.8pt,
            fill=blue!10,
        },
        process/.style={
            draw,
            rectangle,
            align=center,
            font=\large\bfseries,
            line width=0.8pt,
            inner xsep=0.5cm, inner ysep=0.5cm,
            fill=green!15,
        },
        arrow/.style={->, line width=0.8pt},
        line/.style={-, line width=0.8pt},
        dottedline/.style={-, line width=0.6pt, dashed},
        dottedarrow/.style={->, line width=0.6pt, dashed},
    ]
    \node[process](parsing) {\textbf{Start Parsing}};
    \node[process, below=of parsing](if_statement) {\textbf{If Statement}};
    \node[process, below=of if_statement] (if_node1) {\textbf{Create ast.If node}};
    \node[process, below=of if_node1](push1) {\textbf{Push to control stack}};

    \node[state, below=of push1] (state1) {Next Statement?};

    \draw[arrow] (parsing) -- (if_statement);
    \draw[arrow] (if_statement) -- (if_node1);
    \draw[arrow] (if_node1) -- (push1);
    \draw[arrow] (push1) -- (state1);

    \node[process, left=4.5cm of state1] (else_if) {\textbf{Else If Statement}};

    \node[process, below=2cm of state1.south, xshift=-4cm] (else) {\textbf{Else Statement}};

    \node[process, below=2cm of state1.south, xshift=4cm,
          text width=4cm, align=center] (append_prev)
    {\textbf{Append to previous element\\of the control stack}};

    \node[process, right=4.5cm of state1] (end_if) {\textbf{End If Statement}};

    \draw[arrow] (state1) -- node[above,font=\normalsize]{elif} (else_if);
    \draw[arrow] (state1) -- node[above,font=\normalsize]{no more branches} (end_if.west);
    \draw[arrow] (state1.south west) -- node[left,font=\normalsize]{else} (else.north);
    \draw[arrow] (state1.south east) -- node[right,font=\normalsize]{other stmt} (append_prev.north);
    \draw[arrow] (append_prev.north) to[out=100,in=-20] (state1.south east);

    \node[process, below=8cm of else_if] (new_if_node1) {\textbf{Create new ast.If node}};
    \node[process, below=of else] (push_orelse) {Push 'orelse' list of previous \\if to control stack};
    \node[state, below=7cm of end_if] (state2) {Elif counter \textgreater 0?};

    \draw[arrow] (else_if) -- (new_if_node1);
    \draw[arrow] (else) -- (push_orelse);
    \draw[arrow] (end_if) -- (state2);
    \draw[arrow] (push_orelse.east)
        -| (state1.south);

    \node[process, below=of new_if_node1] (append_orelse) {\textbf{Append to orelse of previous if}};
    \node[process, below=of append_orelse] (push_cs) {\textbf{Push to control stack}};

    \draw[arrow] (new_if_node1) -- (append_orelse);
    \draw[arrow] (append_orelse) -- (push_cs);

    \node[process, below=1cm of state2.south, xshift=-3.8cm] (pop_cs) {\textbf{Pop from control stack}};
    \node[process, below=1cm of state2.south, xshift=3.8cm] (pop_if_cs) {\textbf{Pop IF from control stack}};

    \node[process, below=of pop_cs] (decrement) {\textbf{Decrement elif counter}};
    \node[terminal, below=of pop_if_cs] (done) {\textbf{Done}};

    \draw[arrow] (state2.south west) -- node[left,font=\normalsize]{yes} (pop_cs);
    \draw[arrow] (state2.south east) -- node[right,font=\normalsize]{no} (pop_if_cs);
    \draw[arrow] (pop_cs) -- (decrement);
    \draw[arrow] (pop_if_cs) -- (done);
    \draw[arrow] (decrement.east) -| (state2.south);

    \node[process, below=2cm of push_cs] (increment) {\textbf{Increment elif counter}};
    \draw[arrow] (push_cs) -- (increment);
    \draw[arrow] (increment.east) -| (state1.south);

\end{tikzpicture}

    \caption{Internal diagram of ELSE/ELSE IF in Python AST with respective to Contorl Stack}
    \label{fig:if_control}
\end{figure}
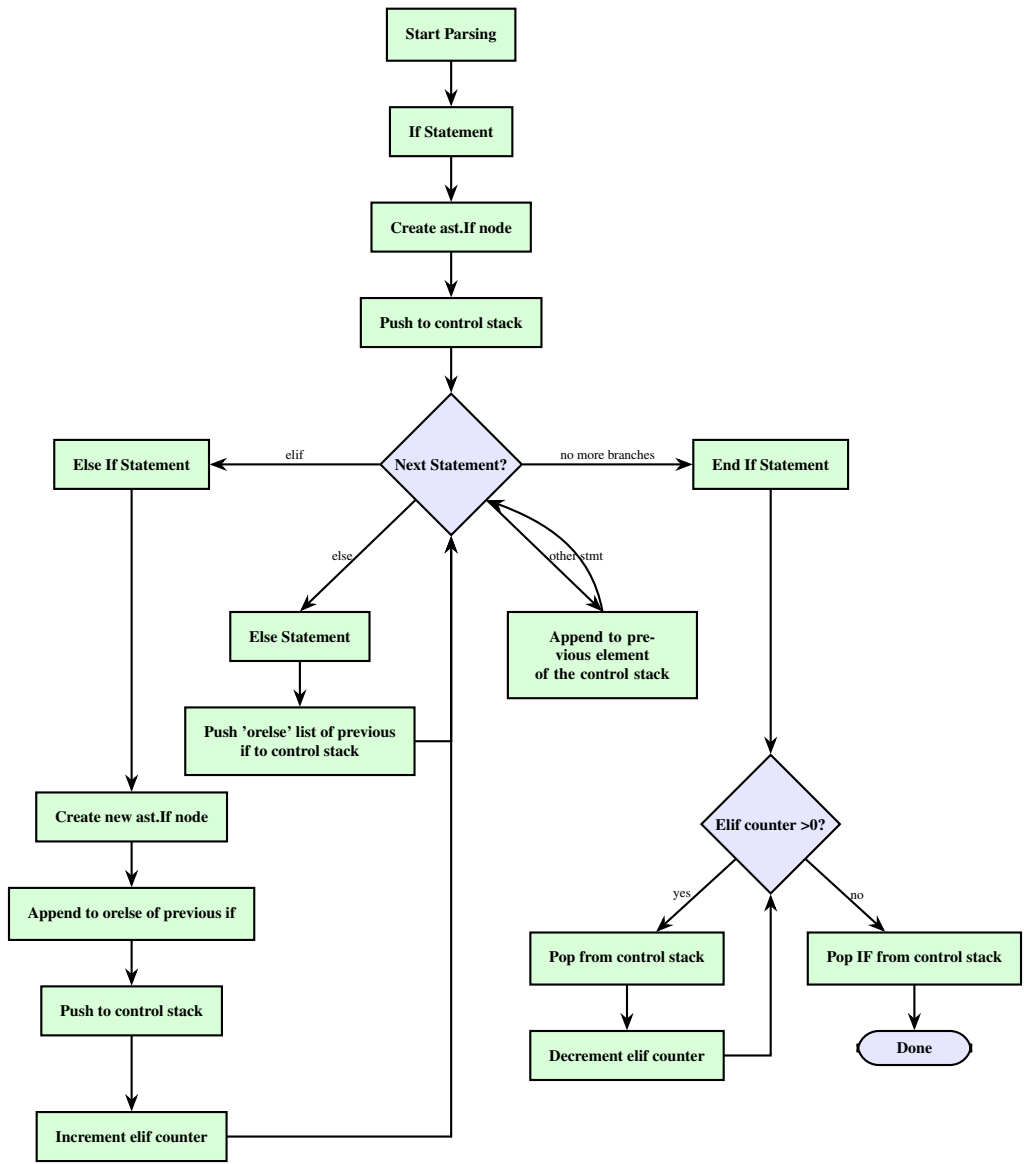

\end{appendix}


\begin{thebibliography}{}


\bibitem[Bradbury et al.(2018)]{jax2018github}
Bradbury J., Frostig R., Hawkins P., Johnson M. J., Katariya Y., Leary C., Maclaurin D., Necula G., Paszke A., VanderPlas J., Wanderman-Milne S., Zhang Q. (2018) {JAX}: composable transformations of {P}ython+{N}um{P}y programs, \url{http://github.com/jax-ml/jax}.

\bibitem[Boucher et al.(2020)]{boucher_2020}
Boucher O., Servonnat J., Albright A. L., Aumont O., Balkanski Y., Bastrikov V., Bekki S., Bonnet R., Bony S., Bopp L., et al. (2020) Presentation and evaluation of the IPSL-CM6A-LR climate model. \textit{Journal of Advances in Modeling Earth Systems}, 12(7), e2019MS002010.

\bibitem[Häfner et al.(2018)]{hafner_veros_2018}
Häfner D., Jacobsen R. L., Eden C., Kristensen M. R. B., Jochum M., Nuterman R., Vinter B. (2018) Veros v0.1 – a fast and versatile ocean simulator in pure Python, \textit{Geoscientific Model Development} \textbf{11}(8), 3299--3312.

\bibitem[Hasco{\"e}t and Pascual(2013)]{hascoet_2013}
Hasco{\"e}t L., Pascual V. (2013) The Tapenade automatic differentiation tool: Principles, model, and specification. \textit{ACM Transactions on Mathematical Software}, 39(3), 1--43.

\bibitem[Kidger and Garcia(2021)]{kidger_equinox_2021}
Kidger P., Garcia C. (2021) Equinox: neural networks in {JAX} via callable {PyTrees} and filtered transformations, \textit{arXiv preprint} arXiv:2111.00254.

\bibitem[Koldunov et al.(2026)]{koldunov_ocean_2026}
Koldunov N. V., Cheedela S. K., Danilov S., Sidorenko D., Beyer S., Jung T. (2026) An {Ocean} {Model} {Ported} by a {Large} {Language} {Model}: {Experience} and {Lessons} from {FESOM2} ({Fortran} to {C} to {C}++/{Kokkos}), \textit{arXiv preprint} arXiv:2606.11356.

\bibitem[Lahlou et al.(2026)]{lahlou_systematic_2026}
Lahlou A., Hawkins L., Gentine P. (2026) Systematic {LLM} {Translation} of {Legacy} {Scientific} {Code} to {Differentiable} {Frameworks}: {Application} to a {Land} {Surface} {Model}, \textit{arXiv preprint} arXiv:2606.07681.

\bibitem[Peterson(2009)]{peterson_2009}
Peterson P. (2009) {F2PY}: a tool for connecting {Fortran} and {Python} programs, \textit{International Journal of Computational Science and Engineering} \textbf{4}(4), 296--305.

\bibitem[STFC(2026)]{fparser_2026}
UKRI Science and Technology Facilities Council (2026) stfc/fparser, \url{https://github.com/stfc/fparser}.

\bibitem[Van Der Walt et al.(2011)]{van_der_walt_numpy_2011}
Van Der Walt S., Colbert S. C., Varoquaux G. (2011) The {NumPy} array: a structure for efficient numerical computation, \textit{arXiv preprint} arXiv:1102.1523.

\bibitem[Zhou et al.(2024)]{zhou_proof_concept_2024}
Zhou A., Hawkins L., Gentine P. (2024) Proof-of-concept: {Using} {ChatGPT} to {Translate} and {Modernize} an {Earth} {System} {Model} from {Fortran} to {Python}/{JAX}, \textit{arXiv preprint} arXiv:2405.00018.

\end{thebibliography}
\end{document}